\documentclass[preprint,showpacs,preprintnumbers,amsmath,amssymb, superscriptaddress,longbibliography,nofootinbib]{revtex4-1}
\usepackage{graphicx} % Required for inserting images
\usepackage{xcolor}
\usepackage{amsmath}
\usepackage{hyperref}
\usepackage[normalem]{ulem}
\usepackage[top = 2cm, bottom = 2cm, right = 2cm, left =2cm]{geometry}

\begin{document}
\title{Five dimensional rotating and Quintessence black hole and their hypershadows}
\author{Milko Estrada }
\email{milko.estrada@gmail.com}
\affiliation{Departamento de Física, Facultad de Ciencias, Universidad de Tarapacá, Casilla 7-D, Arica, Chile}
\author{L. C. N. Santos }
\email{luis.santos@ufsc.br}
\affiliation{ Departamento de Física, CFM - Universidade Federal de Santa Catarina; C.P. 476, CEP 88.040-900, Florianópolis, SC, Brazil }
\author{M. S. Cunha }
\email{marcony.cunha@uece.br }
\affiliation{ Universidade Estadual do Ceará (UECE), Centro de Ciências e Tecnologia, CEP 60.714.903, Fortaleza-CE,  Brazil }

\date{\today}

\begin{abstract}
We present a new five-dimensional rotating quintessence black hole solution. To obtain this, we employ the $5D$ version of the Janis Newman algorithm, which incorporates the Hopf bifurcation. The variation of the quintessence parameter $w_q$ causes the geometry to transition from a regular rotating universe surrounded by a cosmological horizon to a singular rotating geometry, which can represent a naked singularity, a singular extremal black hole, or a singular black hole with both an inner and an outer (event) horizon. We have also determined the properties of the ergosphere. For the study of the shadow, we followed a novel approach in which the $2D$ shadow observed by humans corresponds to cross sections of the $3D$ shadow. We analyzed how quintessence affects both the size and shape of the black hole shadow, showing that increasing the quintessence strength reduces the shadow radius, contrary to the known results in $4D$. We also propose a speculative methodology to test the shadow behavior in five dimensional scenarios, in light of the constraints provided by the Event Horizon Telescope (EHT) concerning the shadow of the four-dimensional supermassive black hole M87. We identify scenarios in which the theoretical $5D$ results could be consistent with these observational constraints. We have also tested the circularity deviation of our shadows, finding that the results satisfy the bound $\Delta C \leq 0.1$ both in the case with quintessence and in the limiting case without quintessence. Finally, we determine the energy conditions required to support the solution.
\end{abstract}

\maketitle

\section{Introduction}

In recent decades, the study of black holes in the presence of extra dimensions has been extensively researched. In this regard, references \cite{Cai:2020igv,LIGOScientific:2016kms,LIGOScientific:2017vwq} suggest that at high energy scales of 1 TeV, the production of black holes in higher dimensions at future colliders becomes a conceivable possibility. Additionally, the AdS/CFT correspondence \cite{Maldacena:1997re} has also attracted attention in the study of higher-dimensional black holes. In particular, a weakly coupled gravitational theory in five-dimensional AdS spacetime corresponds to a strongly coupled gauge field theory in four dimensions \cite{Cai:2020igv}.

On the other hand, recent observations have revealed the existence of gravitational waves generated by the collision of rotating black holes \cite{LIGOScientific:2016aoc}. Thus, both the theoretical and observational studies of rotating black holes have garnered significant attention in recent years. It is worth noting that references \cite{Shaymatov:2018fmp,Shaymatov:2020tna} indicate that the solution of a rotating black hole in five dimensions has an interesting feature: it could be overspun under linear accretion, meaning that it could be converted into a naked singularity by overcharging or spinning, thereby violating the Cosmic Censorship Conjecture, when it has two rotations, but not when it has only one rotation. Thus, both due to the physical interest associated with the study of extradimensional black holes mentioned above, and the physical interest in rotating black holes, the study of rotating black holes in five dimensions is of significant physical relevance.

It is well known that, to generate rotating black hole solutions in four dimensions, the Janis–Newman algorithm is typically applied to a static, spherically symmetric four-dimensional solution. However, in the five-dimensional case, the equations of motion for the rotating scenario become highly nonlinear, making the generation of rotating solutions nontrivial. Until reference \cite{Erbin:2014lwa} provided a five-dimensional version of this algorithm, the number of rotating solutions in five dimensions was quite limited. Remarkably, this reference \cite{Erbin:2014lwa} noted that by rewriting the angular components of the cross-section of a five-dimensional static black hole (BH), corresponding to an $S^3$ sphere, using Hopf bifurcation, it is possible to derive a five-dimensional version of the Janis–Newman algorithm. Consequently, by applying this algorithm to static five-dimensional solutions, rotating solutions can be obtained. See other approach in reference \cite{Tavakoli:2020uzr}.

The geometries of rotating five-dimensional black holes, using Hopf coordinates, have been studied in various contexts. Some examples include the physical characteristics of the Kerr-Myers vacuum case \cite{Myers:1986un,Sakaguchi:2005mz,Erbin:2014lwa, Mirzaiyan:2017adt,Nadi:2019nmi}, regular black holes (RBHs) sourced by nonlinear electrodynamics \cite{Ahmed:2020jic,Sharif:2021vex}, electrically charged Bardeen RBHs \cite{Amir:2020fpa}, and Einstein–Maxwell–Chern–Simons black holes \cite{Amir:2017slq}. See also \cite{Ali:2024pyl}

On the other hand, one of the most important phenomena in cosmology is the accelerated expansion of the universe. One of the theories that provides an explanation for this phenomenon is dark energy, which is associated with the generation of a negative pressure strong enough to accelerate the expansion of the universe. Although this energy is composed of particles that do not interact or interact weakly, it is associated with an amount of energy that alters the geometry of spacetime. For example, dark energy can deflect light from distant stars. In this way, it is of physical interest to understand the influence of dark energy on the properties of objects such as black holes. There are several candidates for dark energy. One of them is the well-known cosmological constant, and another is Quintessence. Quintessence is characterized by a state parameter $w_q$, which is the ratio of pressure to the energy density of dark energy, and the value of $w_q$ falls within the range $-1 <w_q < 0$ in the four-dimensional case. In reference \cite{Kiselev:2002dx}, Kiselev was the first to derive the solutions of a black hole surrounded by quintessence. It is also worth noting that four-dimensional rotating black holes with quintessence models have been studied in the literature \cite{Ghosh:2015ovj}.

On the other hand, one of the most significant achievements of the Event Horizon Telescope (EHT) collaboration has been the capture of images of the shadows of the supermassive black holes M87 \cite{EventHorizonTelescope:2021srq} and Sgr A \cite{EventHorizonTelescope:2022xqj}. In this context, the study of shadow behavior in scenarios involving extra dimensions has also gained attention: for example, reference \cite{Nozari:2024jiz} addresses Scalar-Tensor-Vector theories,and reference \cite{Paithankar:2023ofw}, where connections are established between the parameters of black hole solutions and the structure of the shadow in extradimensional and asymptotically flat scenarios for Pure Lovelock and $f(R)$ gravity. See also \cite{Vagnozzi:2019apd}, where constraints on extra-dimensional spacetimes are derived using data from M87, and \cite{Vagnozzi:2022moj}, where shadow properties are studied using data from Sgr A. The methodology for studying the equations of motion of null geodesics in $5D$, in Hopf coordinates, was initially proposed in reference \cite{Frolov:2003en} for the rotating vacuum solution of Myers-Perry. This methodology is based on the Hamilton-Jacobi formalism. This formalism has since been used for the study of shadow structures in $5D$ in these coordinates. For example, see reference \cite{Papnoi:2014aaa}, where the shadow of the $5D$ Myers-Perry solution is studied. See posterior applications in references \cite{Ahmed:2020jic,Amir:2017slq}.

Due to the fact that the Janis-Newman algorithm in $5D$ requires the use of Hopf coordinates, the study of shadows in $5D$ also requires a different approach compared to the $4D$ case. Typically, in $5D$, shadows have been analyzed, in broad terms, as a $2D$ shape on an image plane at the observer's location \cite{Papnoi:2014aaa,Ahmed:2020jic,Hertog:2019hfb}. As indicated in reference \cite{Novo:2024wyn}, this approach is inspired by the way humans perceive images through the projection of light rays onto our retina, which is a $2D$ surface. Recently, in the same reference \cite{Novo:2024wyn}, a new methodology has been proposed for the study of shadows in higher dimensions: One should adopt the perspective of higher-dimensional beings, whose equivalent of the retina is a volume, meaning that it has $3$ spatial dimensions. Thus, the authors argue that the $2D$ shadow observed by humans would correspond to cross-sections of this $3D$ shadow. In this way, this methodology allows for a more accurate estimation of the behavior and shape of shadows in $5D$. It is worth mentioning that, in our work, this latter methodology will be used for the analysis of shadows.

In this work, we provide a new five-dimensional rotating black hole solution surrounded by quintessence by applying the $5D$ version of the Janis-Newman algorithm, which considers Hopf coordinates, to the static, spherically symmetric quintessence $5D$ black hole solution \cite{Chen:2008ra}. We will study its physical properties, such as the structure of the horizons and the ergosphere. We analyze under what parameters the geometry represents a rotating black hole, a naked singularity, or a rotating geometry surrounded by a cosmological horizon. We determine the behavior of the energy conditions. Additionally, following the methodology for studying shadows in $5D$ recently formulated in reference \cite{Novo:2024wyn}, and described in the previous paragraph, we study the behavior of the shadows and their physical properties for our case. We also discuss our findings in relation to other studies of higher-dimensional black holes in the presence of dark energy, as well as four-dimensional black hole models with quintessence. Furthermore, we provide a speculative methodology to test the shadow behavior in five-dimensional scenarios, in light of the constraints provided by the Event Horizon Telescope (EHT) regarding the shadow size of the four-dimensional supermassive black hole M87.

\section{A new $5D$ rotating and spherically symmetric black hole solution with quintessence}

We will begin by writing the line element of a $5D$ static black hole
\begin{equation} \label{Estatico}
   ds^2= -f(r) \, dt^2 + \frac{dr^2}{f(r)} + r^2 \, d\Omega_3^2
\end{equation}
where the cross-section is written in Hopf coordinates:
\begin{equation} \label{Rotante1}
  d\Omega_3^2 = d\theta^2 + \sin^2 \theta \, d\phi^2 + \cos^2 \theta \, d\psi^2
\end{equation}
where $0 \leq \theta \leq \frac{\pi}{2}$ and $0 \leq \phi, \psi \leq 2\pi$. We can see a review of the Hopf coordinates in references \cite{Sakaguchi:2005mz,Erbin:2014lwa}. 

We use the following ansatz
\begin{equation}
    f(r) = 1-m(r)/r^2
\end{equation}

In our case, to obtain our $5D$ rotating and spherically symmetric Quintessence black hole, we apply the $5D$ Janis-Newman algorithm \cite{Erbin:2014lwa} to the static line element \eqref{Estatico}, which, in our case, corresponds to the five-dimensional static and spherically symmetric Quintessence black hole studied in reference \cite{Chen:2008ra}. See Appendix \ref{RevisionKiselev}. The mass function $m(r)$ corresponds to

\begin{equation} \label{FuncionDeMasa}
    m(r)=2M+ \frac{c}{r^{4w_q}}
\end{equation}
and the line element is
\begin{equation} \label{ElementoDeLineaRotante}
    ds^2 = -dt^2 + \frac{m(r)}{\rho^2} (dt - \varpi)^2 
+ \frac{r^2 \rho^2}{\Delta} \, dr^2 
+ \rho^2 \, d\theta^2 
+ \left( r^2 + a^2 \right) \sin^2 \theta \, d\phi^2 
+ \left( r^2 + b^2 \right) \cos ^2 \theta d\psi^2
\end{equation}
where:
\begin{align}
    \varpi =& a \sin^2 \theta \, d\phi + b \cos^2 \theta d\psi^2 \\
    \rho^2 =& r^2 + a^2 \cos^2 \theta + b^2 \sin^2 \theta \label{rho2} \\
    \Delta =& \left( r^2 + a^2 \right) \left( r^2 + b^2 \right) - m(r)r^2 \label{Delta1}
\end{align}

The metric \eqref{Rotante1} has three Killing vector fields, 
$\left\{\frac{\partial}{\partial t}, \frac{\partial}{\partial \phi}, \frac{\partial}{\partial \psi}\right\}$, 
where the parameters $a$ and $b$ are related to the conserved quantities associated with
$\frac{\partial}{\partial \phi}$ and $\frac{\partial}{\partial \psi}$, respectively \cite{Sakaguchi:2005mz}.

\subsection{Singularity behavior}
It is direct to check that the Ricci scalar can be written as follow:
\begin{equation}
    R=  \left ( \frac{2m'}{r}+ m'' \right) \cdot \frac{1}{\rho^2}
\end{equation}
where $\rho^2$ is given by equation \eqref{rho2}. 

Now, replacing our mass function \eqref{FuncionDeMasa}
\begin{equation}
    R= \frac{4cw_q(-1+4w_q)}{r^{2(1+2w_q)}} \cdot \frac{1}{\rho^2}
\end{equation}
First, the first term of the left should have a divergence at $r=0$ for $1+2w_q>0 \Rightarrow w_q>-1/2 \Rightarrow w_q \in [-0.5,0]$. 

Secondly, we can observe that there could be another potential divergence in the value of the Ricci scalar at $\rho^2 = 0$: 

\begin{equation}
    \rho^2 = r^2 + a^2 \cos^2 \theta + b^2 \sin^2 \theta=0
\end{equation}

First, we can notice a divergence for the case where $a = 0$, $b \neq 0$, $\theta = \frac{\pi}{2}$ or $b = 0$, $a \neq 0$, $\theta = 0$. Since in $5D$ the rotating solution is constructed using the Janis-Newman algorithm, which for this number of dimensions is represented in Hopf coordinates, it becomes cumbersome to recreate an analogous procedure to Kerr-Schild in usual spherical coordinates, in order to determine if this singularity could give rise to a ring in Cartesian coordinates, as happens in the $4D$ case. What is mentioned in this paragraph is beyond the scope of this work and could be addressed in the future. Therefore, for simplicity, in this work we will limit ourselves to studying the case where $a \neq 0$ and $b \neq 0$.

Thus, for $a \neq 0$ and $b \neq 0$, given the angular variable domain $\theta \in [0, \pi/2]$, there are no divergences such that $\rho^2 = 0$. In this way, the only singularity identified is the one mentioned previously, that is, at $r = 0$ for $w_q \in [-0.5, 0]$. In the following, we will discuss the physical scenarios in which different values of $w_q$ lead.

On the other hand, the five dimensional rotating Kretschmann scalar \cite{Amir:2020fpa} involves a large number of terms when we substitute our mass function into it. However, it is straightforward to check that by doing this, one obtains a singularity analysis similar to the one developed in this subsection for the Ricci scalar.

\subsection{Horizon structure}

In this section, we discuss the horizons associated with the geometry obtained in this work. Horizons can be determined by the condition $g^{rr} = \Delta = 0$. Note that this expression determines a type of coordinate singularity, which can be removed with an appropriate coordinate transformation. This is different from the physical singularities discussed in the previous subsection, which can be localized inside horizons or may represent naked singularities. Thus, by using the definition of $\Delta$, the horizon condition implies the following equation
\begin{equation}
    \Delta = \left( r_*^2 + a^2 \right) \left( r_*^2 + b^2 \right) - (2M + \frac{c}{r_*^{4w_q}})r_*^2 = 0.
\end{equation}
From the previous equation, we can deduce that the mass parameter $M$ such that $\Delta = 0$ is:
\begin{equation} \label{ParametroDeMasa}
   M= -\frac{c r_*^{2 - 4 w_q} - (a^2 + r_*^2) (b^2 + r_*^2)}{2 r_*^2}
\end{equation}
where $r_*$ represents each value of the radial coordinate where $\Delta = 0$ for a fixed value of $M$.

\begin{figure}[ht]
  \begin{center}
          \includegraphics[width=5.1in]{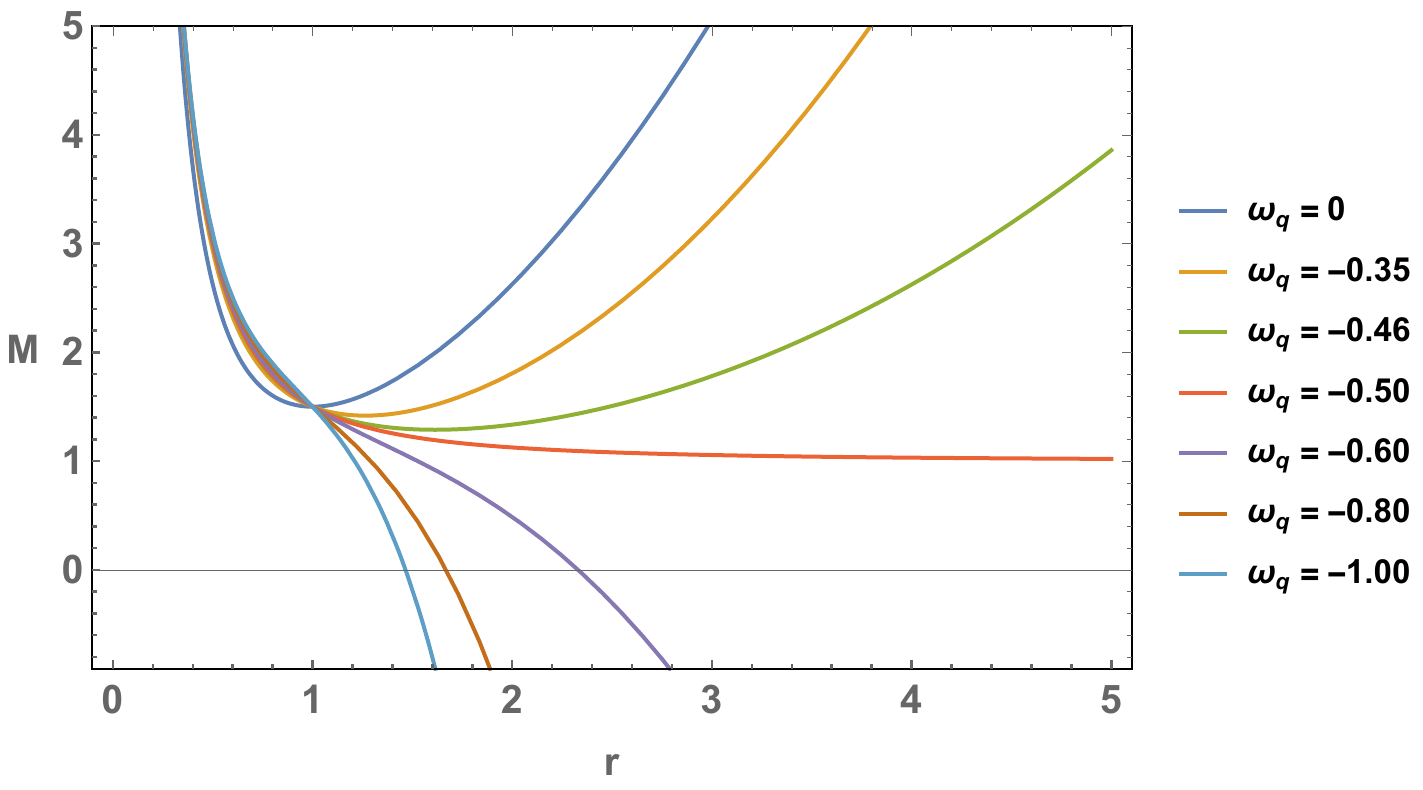}
      \caption{On the horizontal axis, values of $r_*$ such that $\Delta(r_*) = 0$. On the vertical axis, values of the parameter $M$ for $w_q=0,-0.35,-0.46,-0.5,-0.6,-0.8,-1$. Here we use $a=b=c=1$.} \label{MvsW}
  \end{center}
\end{figure}

First, we will analyze how the behavior of the mass parameter $M$ varies for different values of the parameter $w_q$. In this work, for simplicity, we will consider the case where $a = b$ for the analysis of the horizon and ergosphere. In Figure \ref{MvsW}, we notice that for the chosen parameters, when $w_q > w_q^{\text{(crit)}} \sim -0.5$, there can be up to two horizons. The values on the horizontal axis in the descending part of the curves (where $w_q = 0, -0.35, -0.46$) represent the inner horizon, while those in the ascending parts represent the values of the black hole horizon.

\begin{figure}[ht]
  \begin{center}
    \includegraphics[width=5.1in]{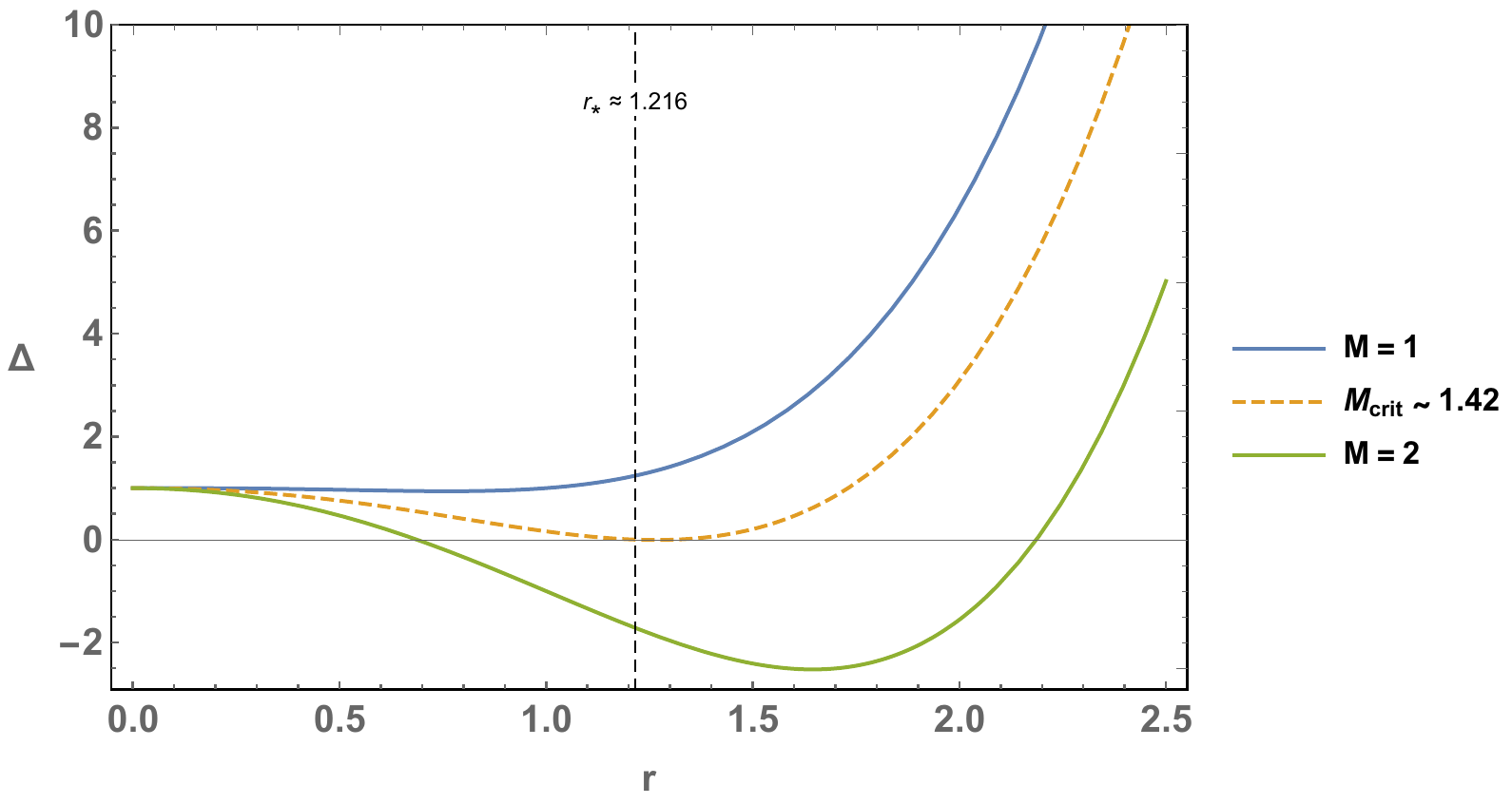}
  \caption{On the horizontal axis, values of $r$. On the vertical axis, the behavior of $\Delta(r)$ for $w_q = - 0.35$. The orange dashed curve represents the extremal black hole case ($M = M_{crit} \simeq 1.42$), where the inner and outer horizons merge ($r_{crit} \simeq 1.216$). For $M>M_{crit}$ (green curve), both the inner and event horizon are present. When $M < M_{crit}$ (blue curve), the absence of horizons indicates a naked singularity. Here we use $a=b=c=1$.}.  \label{DeltavsW}
  \end{center}
\end{figure}

\begin{figure}[ht]
\includegraphics[scale=1]{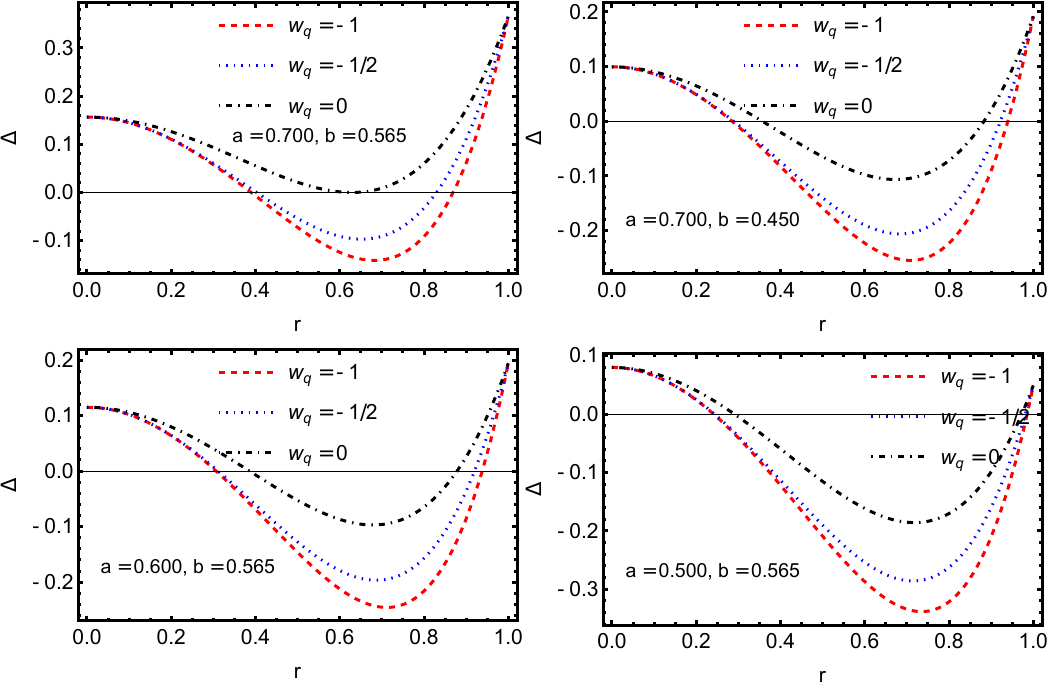}\newline
\caption{This figure presents four plots of \(\Delta\) as a function of \(r\), the radial coordinate of the black hole. The horizons are determined by the points where \(\Delta = 0\), and the number of points where \(\Delta = 0\) indicates the number of horizons.
Each plot corresponds to specific values of the parameters \(a\) and \(b\), as indicated in the respective panels. The curves represent different values of the equation-of-state parameter \(w_q\), with \(w_q = -1\) (red dashed line), \(w_q = -1/2\) (blue dotted line), and \(w_q = 0\) (black dash-dotted line). }
\label{fig1}
\end{figure}

An example of the behavior where $w_q > w_q^{(crit)}$, i.e., where up to two horizons can exist depending on the value of the parameter $M$, can be seen in Figure \ref{DeltavsW}. The orange curve shows the case where $M = M_{crit}$, i.e., where the inner and outer horizons coincide. This case corresponds to a singular extremal black hole. For $M > M_{crit}$, the green curve, there is both an inner horizon and a black hole horizon. Thus, this case represents a singular black hole with the presence of both mentioned horizons. In this regard, see also Figure \ref{fig1}, which shows the behavior of $\Delta$ for values different from the involved parameters.
For $M < M_{crit}$, the blue curve of Figure \ref{DeltavsW}, we observe that there are no horizons, indicating the presence of a rotating $5D$ naked singularity.

\begin{figure}[ht]
  \begin{center}
       \includegraphics[width=5.1in]{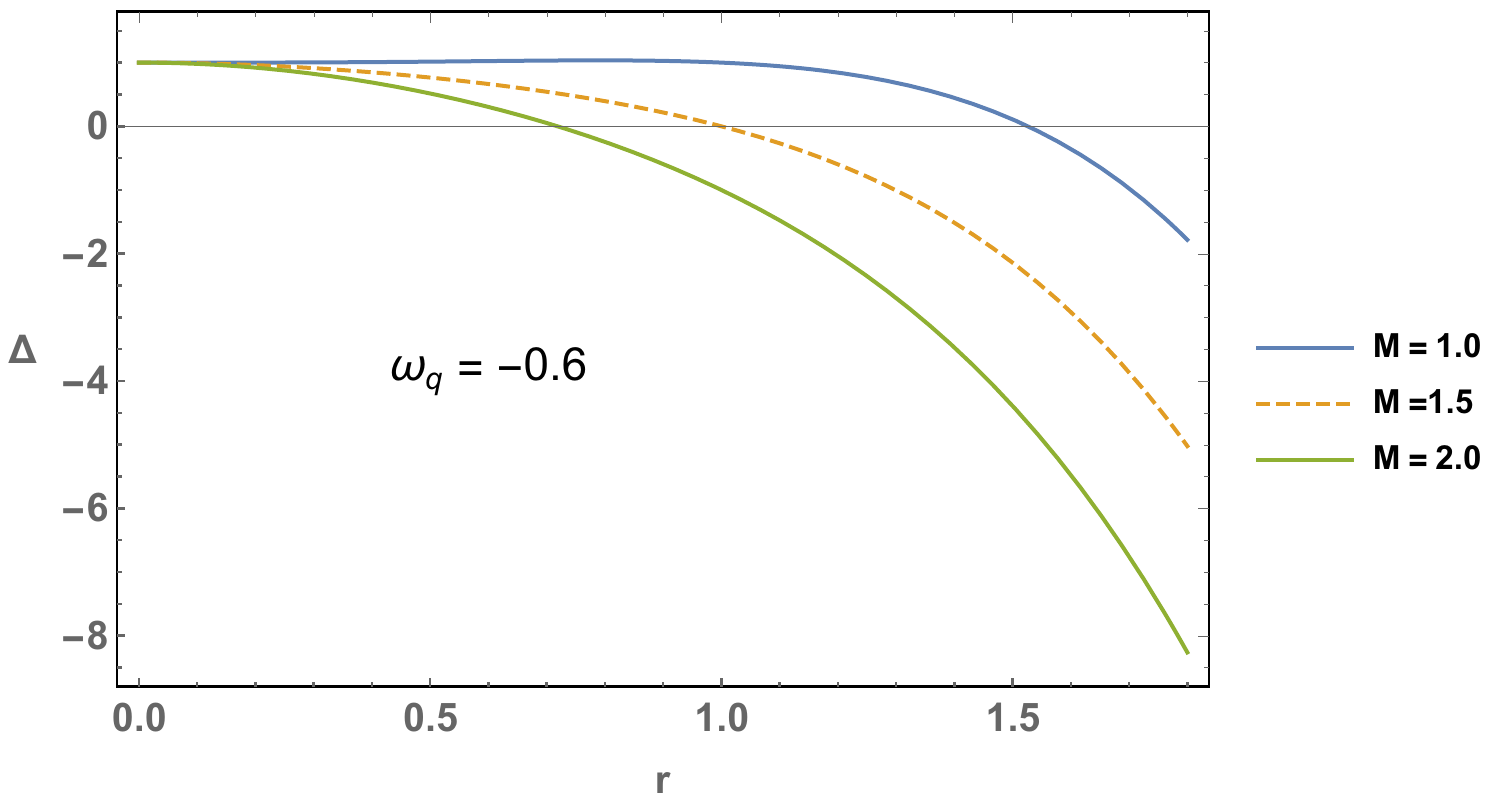}
  \caption{Behavior of $\Delta(r)$ for $w_q = - 0.6$, and different values of $M$. Here, $a=b=c=1$}.\label{DeltavsW1}
  \end{center}
\end{figure}

On the other hand, in Figure \ref{MvsW}, we also notice that for $w_q < w_q^{\text{(crit)}} \sim -0.5$, the mass parameter is always decreasing, meaning there is only one horizon. See the curves with $w_q = -0.6, -0.8, -1$. Thus, in these cases, the rotating geometry possesses a cosmological horizon instead of a black hole horizon. We can see in Figure \ref{DeltavsW1} that for $w_q = -0.6$, regardless of the value of $M$, after crossing the horizon from left to right, the signature of $\Delta$ becomes negative, which describes a cosmological horizon, analogous to what occurs in a de Sitter universe. Therefore, the geometry can be interpreted as a kind of spherically symmetric and rotating universe. For the chosen parameters, there is no central singularity, since for $w_q < -0.5$, the geometry is regular according to the analysis in the previous subsection. 

In other words, the variation (increase) of the quintessence parameter causes the geometry to transition from a regular rotating universe surrounded by a cosmological horizon (for $w_q < w_q^{\text{(crit)}}$) to a singular rotating geometry (once the parameter reaches a critical value $w_q^{(crit)}$), which can represent (depending on the value of the mass parameter $M$) a naked singularity, a singular extremal black hole, or a singular black hole with an inner horizon and an event horizon.

The static limit surface is another relevant physical  feature to characterize the obtained geometry. This surface defines the outer boundary of the ergosphere, where the frame-dragging becomes inevitable. It is obtained from the temporal component of spacetime metric demanding that $g_{tt} = 0$, which implies the expression
\begin{equation} \label{Ergosfera}
 2M+ \frac{c}{r^{4w_q}} - (r^2 + a^2 \cos^2 \theta + b^2 \sin^2 \theta) = 0.   
\end{equation}

\begin{figure}[ht]
\includegraphics[scale=0.55]{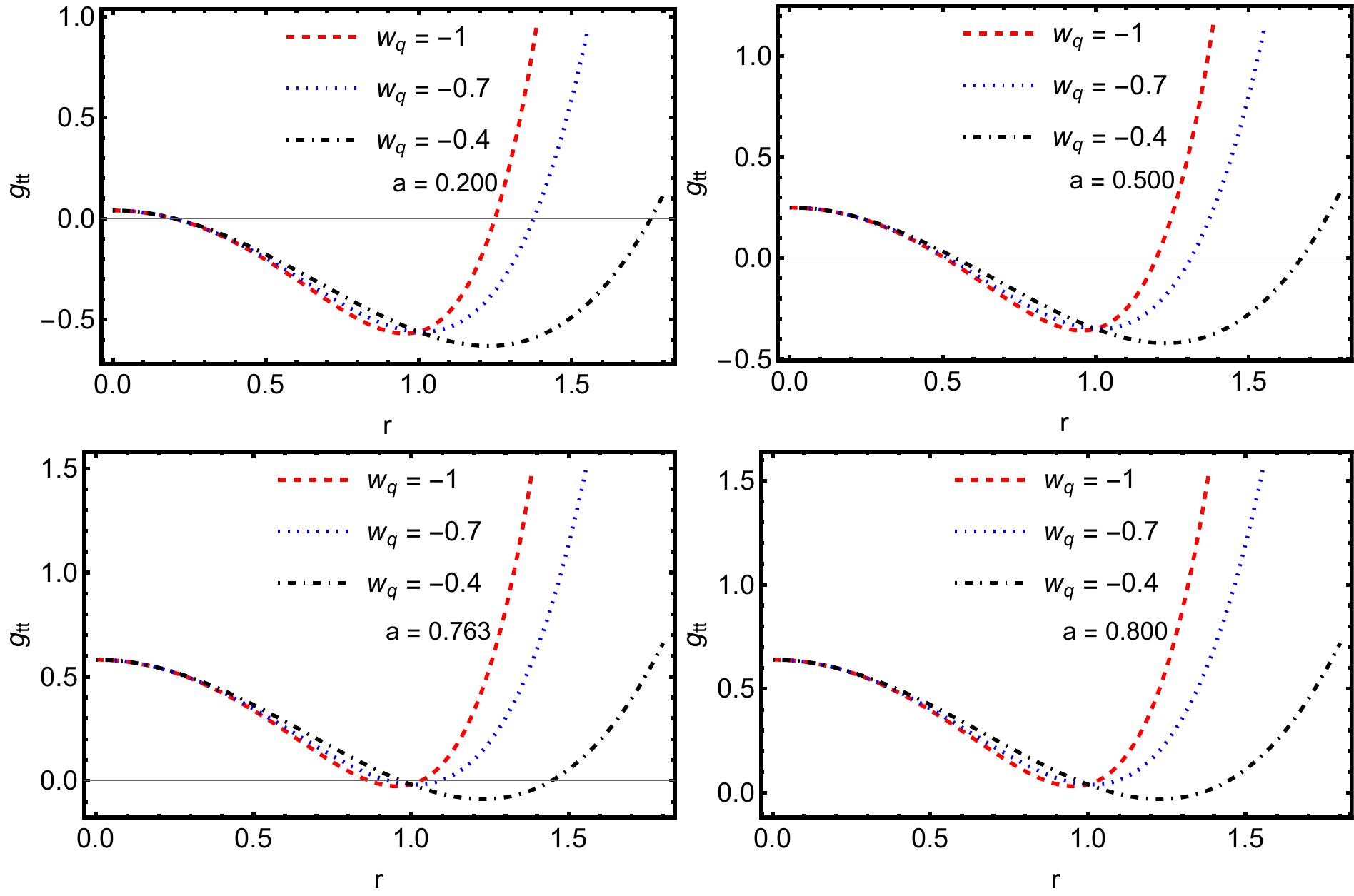}\newline
\caption{The figure shows the behavior of static limit surface for the rotating and spherically symmetric quintessence black hole for four different values of $a$ and considering $b=0$ and $c=-0.4$. Each graph has 3 curves with varying values of $w_q$.}
\label{fig2}
\end{figure}

Figure \ref{fig2} shows the static limit for different values of parameter values of our solution. We consider four values for the parameter $a$. As we can see, increasing values of $a$ tend to decrease the distance between the roots of each curve and for a limiting value of $a$ there are no roots in the geometry. 

\begin{figure}[ht]
  \begin{center}
      \includegraphics[width=6in]{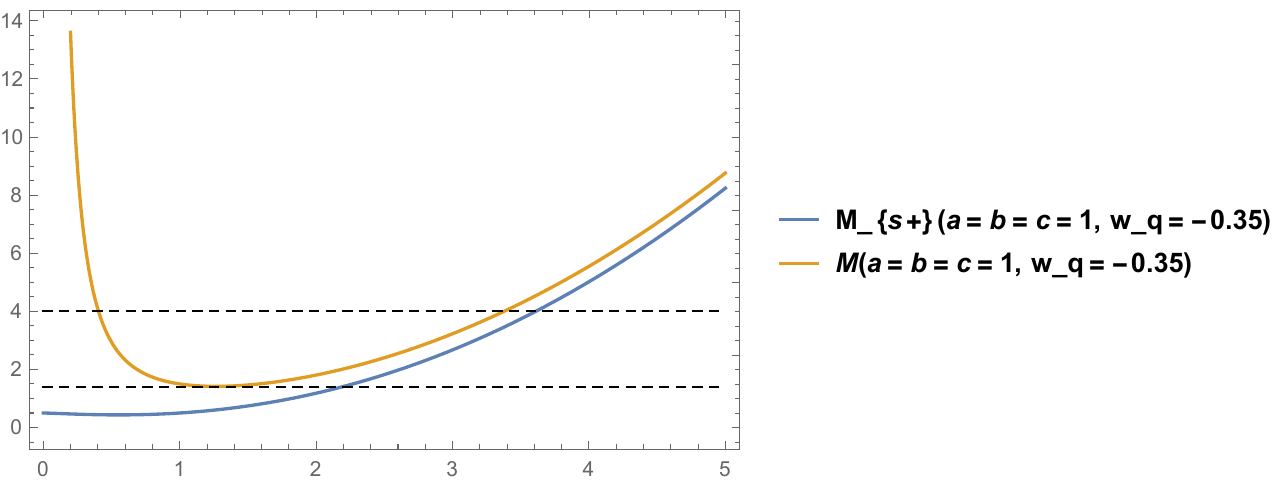}
        \caption{On the horizontal axis, values of $r=r_*$ such that $\Delta=0$ and $g_{tt}=0$. On the vertical axis, values of $M$ for $\Delta=0$ and $M_{s+}$ for $g_{tt}=0$}\label{FigMErgosfera}
  \end{center}
\end{figure}

As previously stated, we restrict our analysis to the case $a = b$, i.e. also in the analysis of the horizon and ergosphere. In this case, the parameter $M$, obtained from equation \eqref{Ergosfera}, will be referred to as $M_{s+}$ to distinguish it from the one obtained from equation \eqref{ParametroDeMasa} described earlier ($M$). Despite the aforementioned, it is worth mentioning that $M_{s+} = M$, meaning they have the same value, as both represent the same mass parameter of the black hole solution. Thus, we have
\begin{equation}
    M_{s+}= \frac{a^2+r^2}{2}- \frac{c}{2r^{4w_q}}
\end{equation}
%%%%
In Figure \ref{FigMErgosfera}, we observe the behavior of the mass parameters $M_{s+}$ and $M$. It is important to mention that for the analysis to be valid, the condition $M_{s+}=M$ must be satisfied, as both are associated with the unique mass of the black hole. For example, in the upper segmented curve at $M_{s+}=M=4$, we observe the presence of an inner horizon and the black hole horizon in the descending and ascending parts of the orange curve. We also notice that, as expected, the black hole horizon is covered by the point $r_{s+}$, which corresponds to the interaction of the blue curve with the segmented line. The region between $r_+ < r < r_{s+}$, i.e. the ergosphere, is included here. On the other hand, when $M_{s+}=M=M_{crit}$, which, for the parameters in the figure, corresponds to $M_{s+}=M=M_{crit} \sim 1.75$, we observe that the ergosphere corresponds to the region between the extremal black hole horizon and $r_{s+}$.

\begin{figure}[ht]
  \begin{center}
      \includegraphics[width=6in]{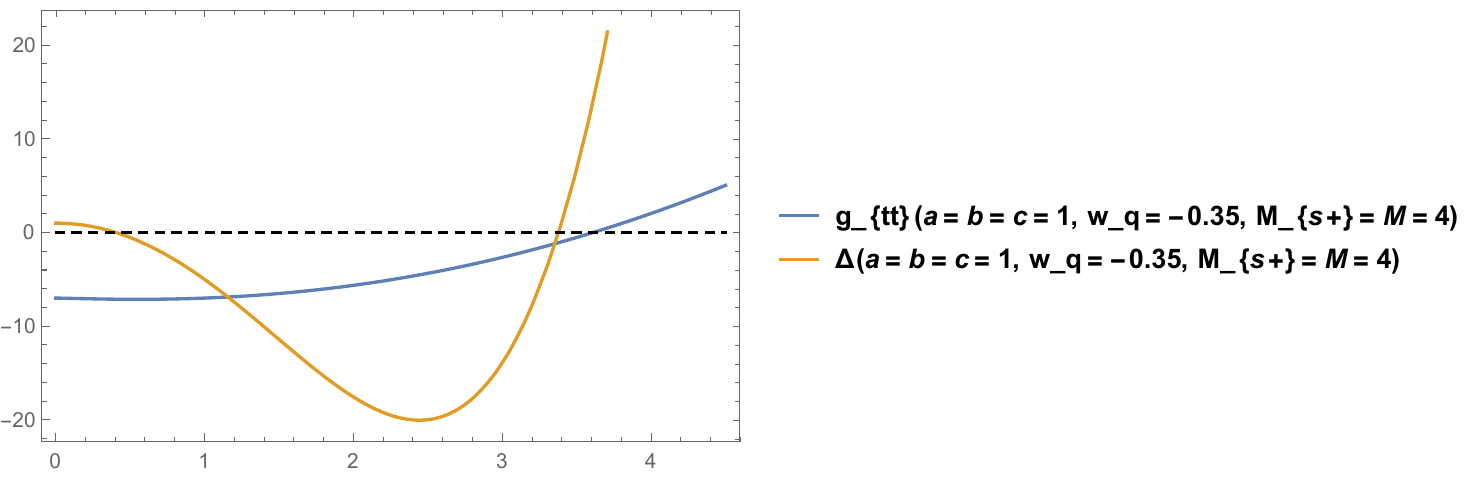}
        \caption{On the horizontal axis, values of radial coordinate $r$. On the vertical axis, values of $\Delta$ and $g_{tt}$}\label{FigGttErgosfera}
  \end{center}
\end{figure}

Thus, in Figure \ref{FigGttErgosfera}, we also observe the previously described behavior for $M_{s+}=M=4 > M_{cri}$, meaning we see an inner horizon, a black hole horizon, and an ergosphere that lies between $r_+ < r < r_{s+}$.

\section{Geodesics and shadows}
As mentioned in reference \cite{Ahmed:2020jic}, a compact object casts a shadow when it is situated between the observer and a source emitting photons (\textit{i.e.}, light). If a black hole is positioned between a bright object, such as quasars, and the observer, it will cast a shadow \cite{Ahmed:2020jic}. The apparent shape of a black hole is defined by the boundary of the shadow. To estimate the shape of a black hole's shadow, we must study the motion of photons.

The methodology for studying the equations of motion of null geodesics in $5D$, in Hopf coordinates, was initially proposed in reference \cite{Frolov:2003en} for the rotating vacuum solution of Myers-Perry. This methodology is based on the Hamilton-Jacobi formalism. See Appendix \ref{ApendiceGeodesicas}. This formalism has been used for the study of shadow structures in $5D$ in these coordinates. For example, see reference \cite{Papnoi:2014aaa}, where the shadow of the $5D$ Myers-Perry solution is studied. See also \cite{Ahmed:2020jic,Amir:2017slq}.

The calculations of the shadow shape are greatly simplified using the variable change $x = r^2$. Thus, in order to simplify the calculations, the mentioned change of variable is used. Following reference \cite{Novo:2024wyn}, this work considers the simplest case where $a=b$. As indicated in this reference, this case possesses enhanced symmetries, and thus the equations of motion simplify greatly. Furthermore, we are interested in a massless photon. From equations \eqref{FuncionDeMasa} and \eqref{Delta1}
\begin{align} 
   & m(x)=2M+ \frac{c}{x^{2w_q}} \label{FuncionDeMasaX} \\
   & \Delta = \left( x + a^2 \right)^2 - m(x)x \label{DeltaX}
\end{align}

On the other hand, by observing the line element \eqref{ElementoDeLineaRotante}, we notice that there are at least three constants of motion. These correspond to the energy, $E$, and the angular momenta in two independent planes defined by the angles $\phi$ and $\psi$
\begin{equation}
    p_t = -E, \quad p_{\phi} = \Phi, \quad p_{\psi} = \Psi.
\end{equation}

We are interested in studying the propagation of photons around the spacetime of the 5D rotating black hole. Specifically, we are interested in observing the shape that the potential existence of the photon sphere takes. From the methodology described in Appendix \ref{ApendiceGeodesicas}, based on equations \eqref{EcuacionAccionJacobianaX} and \eqref{EcuacionX}, for our spacetime, the radial geodesic equation, in the form of the first-order differential equation, is:
\begin{equation} \label{EqMovimientoRadial}
    \rho^4 \dot{x}^2= 4 \mathcal{X}
\end{equation}
where
\begin{equation} \label{MathX}
\mathcal{X}= \Delta \left ( x E^2-K  \right )+ m(x) (x+a^2)^2 \mathcal{E}^2    
\end{equation}
with
\begin{equation} \label{MathE}
   \mathcal{E}=E+ \frac{a}{x+a^2} \left ( \Phi+ \Psi  \right )
\end{equation}
We will follow the methodology proposed in reference \cite{Novo:2024wyn}, where the Carter's constant is redefined such that $K = Q - Ea^2$. Furthermore, the quantity $\alpha$ is defined as $\alpha = \Phi + \Psi$. Furthermore, the scale adjustment $E = 1$ is used, which can be interpreted as a redefinition of the affine parameter along the null geodesics.
\begin{equation} \label{MathX1}
\mathcal{X}= (a^2-Q+x) \left ( (a^2+x)^2-m(x) x  \right ) + \left (m(x) \right )^2 (a^2+a \alpha + x)^2
\end{equation}
For our purposes, that is, the representation of the shadow, we are only interested in null geodesics with a constant radial coordinate. The radial motion is determined by the equation of motion \eqref{EqMovimientoRadial}. Spherical photon orbits have a constant radial coordinate $x$, and therefore they satisfy 
\begin{align}
&\mathcal{X} = 0 \label{Condiciones} \\
& \frac{d\mathcal{X}}{dx} = 0 \nonumber
\end{align}
simultaneously. The usefulness of introducing $\alpha$ now becomes apparent, as this system can be solved for the parameters $Q$ and $\alpha$ along these orbits.

By substituting equation \eqref{MathX1} into condition \eqref{Condiciones}, we obtain the following expression
%%%%%%%%%%%%%%%%%%%%%%%%%%%%%%%%%%%%%%%%%%%%%%%%%%%%%%%%%%%%%%%%%%%%%%%%%%%%%%%%
{\small
\begin{align} \label{FuncionAlfa}
& \alpha(x) = \bigg( a^2 (c - M x^{2 w_q}) \Big( M (-2 a^2 + M - 2 x) x^{1 + 4 w_q} + c^2 (x + 2 w_q x) + 2 c x^{2 w_q} \Big( 2 a^4 w_q +  \nonumber \\
& x \Big( -M (1 + w_q) + x + 2 w_q x \Big) + a^2 (x + 4 w_q x) \Big) \Big) \bigg)^{-1} \cdot \nonumber \\
& \bigg( 4 a^7 c w_q x^{2 w_q} (-c + M x^{2 w_q}) - 
 a^5 x^{1 + 2 w_q} (-c + M x^{2 w_q}) (-c (1 + 12 w_q) + M x^{2 w_q}) - 
 a x^2 (c - M x^{2 w_q}) \cdot \nonumber \\
 & (2 c^2 w_q - M x^{1 + 4 w_q} + c x^{2 w_q} (-2 M w_q + 4 w_q x + x)) - 
 a^3 x (c - M x^{2 w_q}) (c^2 (1 + 2 w_q) + M (M - 2 x) x^{4 w_q} + \nonumber \\
&   2 c x^{2 w_q} (-M (1 + w_q) + x + 6 w_q x)) + \big (   c - 2 c w_q + x^{2 w_q} (2 a^2 - M + 2 x)    \big  )^{-1} \cdot \nonumber \\
& \Big (    a^2 x^{-1 - 4 w_q} (c - M x^{2 w_q}) (c^3 x + c^2 (1 - 3 M + 2 w_q) x^{1 + 2 w_q} - M x^{1 + 6 w_q} (M^2 + 2 a^2 - M + 2 x) + \nonumber \\
& c x^{4 w_q} (4 a^4 w_q + 2 a^2 (4 w_q x + x) + x (3 M^2 - 2 M (1 + w_q) + 2 (1 + 2 w_q) x)))      \Big )^{1/2} \nonumber \\
& \Big ( 2 a^6 x^{1 + 6 w_q} + a^4 x^{1 + 4 w_q} (c - 2 c w_q - (M - 6 x) x^{2 w_q}) + x^{2 + 2 w_q} (c - 2 c w_q - (M - 2 x) x^{2 w_q}) (c + x^{2 w_q} (-M + x)) \nonumber \\
&+ 2 a^2 x^{2 + 4 w_q} (-2 c (-1 + w_q) + x^{2 w_q} (-2 M + 3 x)) \Big )
\end{align}
}
%%%%%%%%%%%%%%%%%%%%%%%%%%%%%%%%%%%%%%%%%%%%%%%%%%%%%%%%%%%%%%%%%%%%%%%%%%%%%%%

%\begin{align} \label{FuncionAlfa}
%& \alpha(x) = \bigg( a^2 \gamma_x \Big( M (-2 a^2 + M - 2 x) x^{1 + 4 w_q} + c^2 \varrho_x + 2 c x^{2 w_q} \Big( 2 a^4 w_q -M x(1 + w_q)+ x\varrho_x  \nonumber \\
%&  + a^2 (x + 4 w_q x) \Big) \Big) \bigg)^{-1} {\color{red}\times} \bigg(-4 a^7 c w_q x^{2 w_q} \gamma_x - a^5 x^{1 + 2 w_q}\gamma_x (-\gamma_x  - 12 c w_q)) - a x^2 \gamma_x {\color{red}\times} \nonumber \\
% & (2 c^2 w_q - M x^{1 + 4 w_q} + c x^{2 w_q} (-2 M w_q + 4 w_q x + x)) - 
% a^3 \gamma_x (c^2 \varrho_x + M (M - 2 x) x^{1+4 w_q} \nonumber \\
%& +  2 c x^{1+2w_q} (-M (1 + w_q) + x + 6 w_q x)) + \big ( \gamma_x - 2 c w_q + x^{2 w_q} (2 a^2 + 2 x)    \big  )^{-1} {\color{red}\times} \nonumber \\
%& \Big (a^2 x^{-1 - 4 w_q} \gamma_x \big(c^3 x + c^2 (\varrho_x - 3 M x) x^{2 w_q} - M x^{1 + 6 w_q} (M^2 + 2 a^2 - M + 2 x) \nonumber \\
%&+ c x^{4 w_q} (4 a^4 w_q + 2 a^2 (4 w_q x + x) + x (3 M^2 - 2 M (1 + w_q) + 2\varrho_x))\big) \Big)^{1/2} \Big ( 2 a^6 x^{1 + 6 w_q} \nonumber \\
%&+a^4 x^{1 + 4 w_q} (\gamma_x - 2 c w_q + 6 x^{1+2 w_q}) + x^{2 + 2 w_q} (\gamma_x - 2 c w_q + 2 x^{2 w_q+1}) (\gamma_x + x^{2 w_q+1}) \nonumber \\
%& + 2 a^2 x^{2 + 4 w_q} \big(2 c (1 - w_q) + x^{2 w_q} (-2 M + 3 x)\big) \Big ),
%\end{align}
%\noindent where $\gamma_x=c- M x^{2 w_q}$, $\varrho_x=x + 2 w_q x$.  

As we will later see graphically, the above expression corresponds to the branch, which is the solution of equation \eqref{Condiciones}, that correctly describes the shape of the shadow, with the remaining branch being of no physical interest.

By substituting the value of $\alpha$ in condition \eqref{Condiciones}, we can obtain the value of $Q$
\begin{equation}
    Q(x) = \frac{(a^2+x) \left (-x^{1-2w_q}c-x^2+(M-2a^2)x-a^4 \right )-(M-cx^{-2w_q})^2 \left ( a (a+\alpha)+x \right )^2}{-(a^2+x)^2+x(M-cx^{-2w_q})}
\end{equation}

\subsection{ About the shape of the $5D$ shadow}

As mentioned, the study of shadows requires a different approach compared to the $4D$ case because the Janiss Newman algorithm in $5D$ requires the use of Hopf coordinates. Typically, in $5D$, shadows have been analyzed, in broad terms, as a $2D$ shape on an image plane at the observer's location \cite{Papnoi:2014aaa,Ahmed:2020jic,Hertog:2019hfb}. As indicated in reference \cite{Novo:2024wyn}, this setup is inspired by the way humans perceive images through the projection of light rays onto our retina, which is a $2D$ surface. Recently, in the same reference \cite{Novo:2024wyn}, it is proposed that: For the study of shadows in five dimensions, one should adopt the perspective of higher-dimensional beings, whose equivalent of the retina is a volume, meaning it has 3 spatial dimensions. Thus, the authors argue that the $2D$ shadow observed by humans would correspond to cross-sections of this $3D$ shadow. The latter is given by the $\bar{X},Y,Z$,  which correspond to the impact parameters of the light rays . It is also assumed that an observer is placed at a point considered to be very far from the black hole. In this way, the authors propose a formalism in which the representation of the shadow is given by the following coordinates:
\begin{align}
&\bar{X} = -\alpha \sin \theta_0 - \cos v \cos \theta_0 \sqrt{Q - \alpha^2} \label{PlotX} ,\\
& Y = -\alpha \cos \theta_0 + \cos v \sin \theta_0 \sqrt{Q - \alpha^2} \label{PlotY}, \\
&Z = \sin v \sqrt{Q - \alpha^2}
\end{align}
where $\theta_0$ corresponds to the observation angle and $v \in [0,2\pi]$ is related with the $Z$ coordinate. In this work, we will analyze the image projected at $Z = 0$, i.e., for $v = 0$.

\begin{figure}[ht]
  \begin{center}
      \includegraphics[width=3.5in]{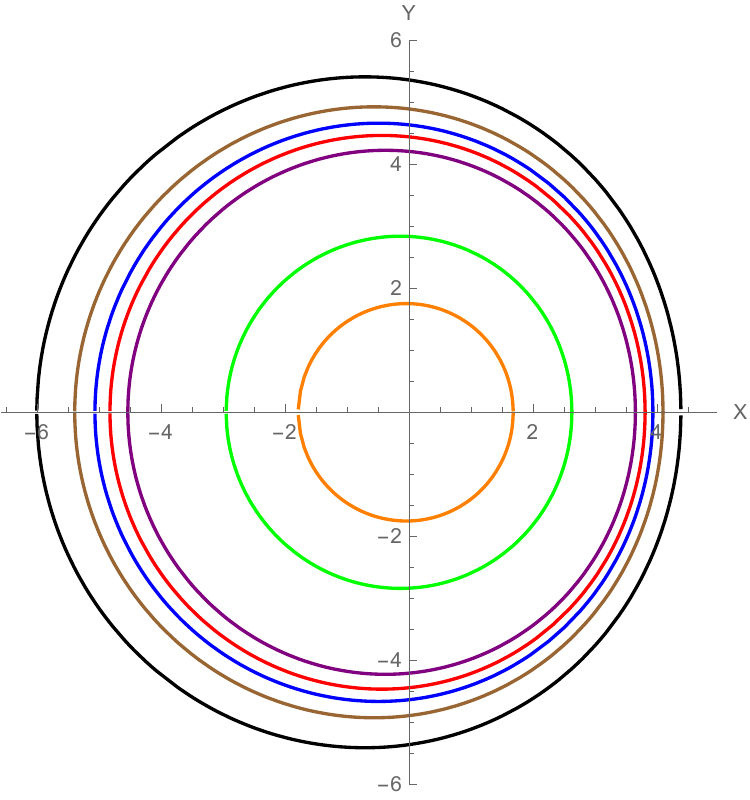}
        \caption{Hypershadow corresponding to a $Z=0$ ($v=0$) slice with $\theta_0=\pi/2,a=0.49,w_q=-0.3,M=4$. In black $c=0.12$, \textcolor{brown}{in brown $c=0.223$}, \textcolor{blue}{in blue $c=0.285$}, \textcolor{red}{in red $c=0.335$}, \textcolor{purple}{in purple $c=0.4$} , \textcolor{green}{in green $c=0.95$}, \textcolor{orange}{in orange $c=2$}.}\label{FigSombras}
  \end{center}
\end{figure}

\begin{figure}[ht]
  \begin{center}
      \includegraphics[width=3.5in]{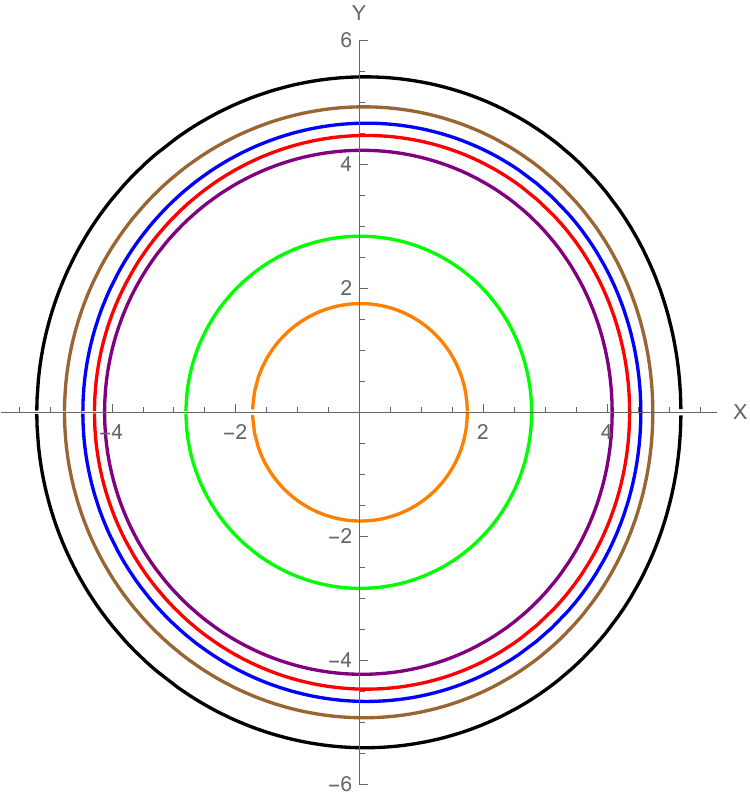}
        \caption{Hypershadow corresponding to a $Z=0$ ($v=0$) slice with $\theta_0=\pi/2,a=0.49,w_q=-0.3,M=4$. In black $c=0.12$, \textcolor{brown}{in brown $c=0.223$}, \textcolor{blue}{in blue $c=0.285$}, \textcolor{red}{in red $c=0.335$}, \textcolor{purple}{in purple $c=0.4$} , \textcolor{green}{in green $c=0.95$}, \textcolor{orange}{in orange $c=2$}.}\label{FigSombras1}
  \end{center}
\end{figure}

The numerical behavior of the shadow is shown in Figure \ref{FigSombras}. In the plots colored black, brown, blue, red, purple, green, and orange, the value of the parameter $c$, corresponding to the quintessence term in equation \eqref{FuncionDeMasa}, increases progressively.
First, we note that the shadow boundary distribution is not symmetric with respect to the vertical axis located at $\bar{X} = 0$. A larger portion of the shadow lies to the left of this axis. In Figure \ref{FigSombras1}, we perform a horizontal translation to the right so that the horizontal radius at $Y = 0$ is the same on both sides of the vertical axis at $\bar{X} = 0$. Thus, we observe that as the magnitude of the quintessence parameter $c$ increases, the size of the shadow decreases. In other words, the presence of quintessence alters the trajectories of photons such that the corresponding photon sphere tends to shrink as the influence of quintessence becomes stronger.

Our result differs from the $4D$ study of the Kerr black hole under the influence of quintessence in reference \cite{Singh:2017xle}. In the latter, if we increase the values of the normalization factor $c$, we find that the size of the black hole shadow increases. This difference may be due to the use of the definition of the observational device in reference \cite{Novo:2024wyn}. The latter assumes that, since spacetime with an extra dimension was considered, observers should also have an extra dimension. Therefore, in our work, as well as in reference \cite{Novo:2024wyn}, we consider the shadow as it would be perceived by a three-dimensional observational device (such as a retina), which means that the shadow is also a three-dimensional object, a volume. It is worth mentioning that in reference \cite{Abdujabbarov:2015pqp}, the rotating case with quintessence influenced by plasma was also studied in $4D$. The authors argue that, with an increase in the plasma refractive index, the apparent radius of the shadow increases. Therefore, it would also be interesting to study the influence of plasma in $5D$ in a future work, using the methodology employed in this study.

\begin{figure}[ht]
  \begin{center}
      \includegraphics[width=2.8in]{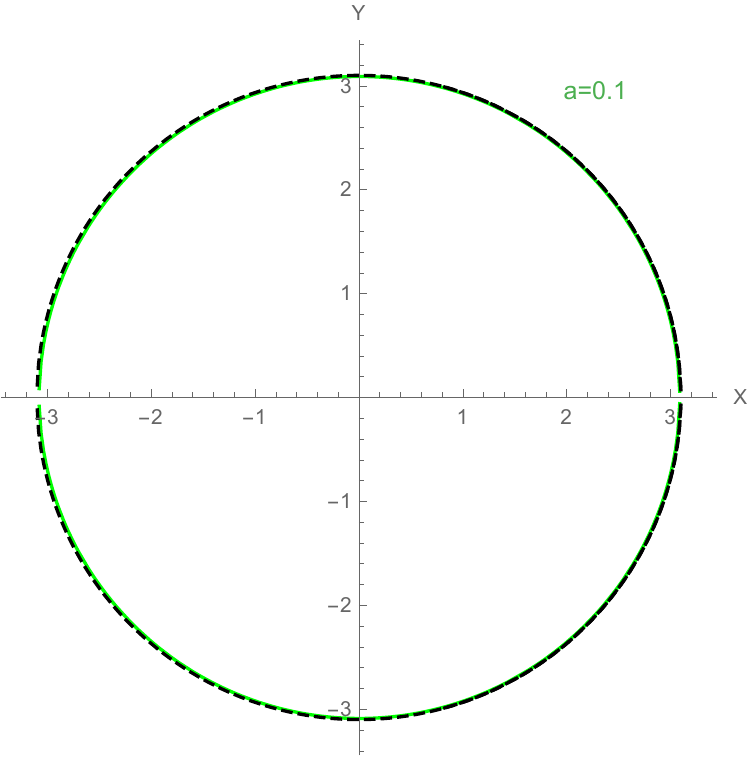}
       \includegraphics[width=2.8in]{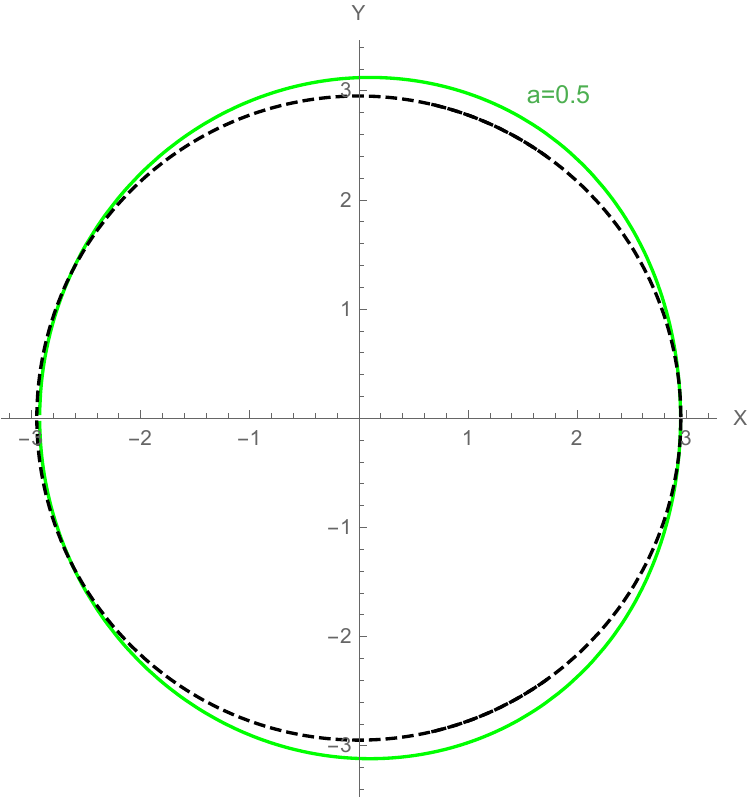}
       \includegraphics[width=2.8in]{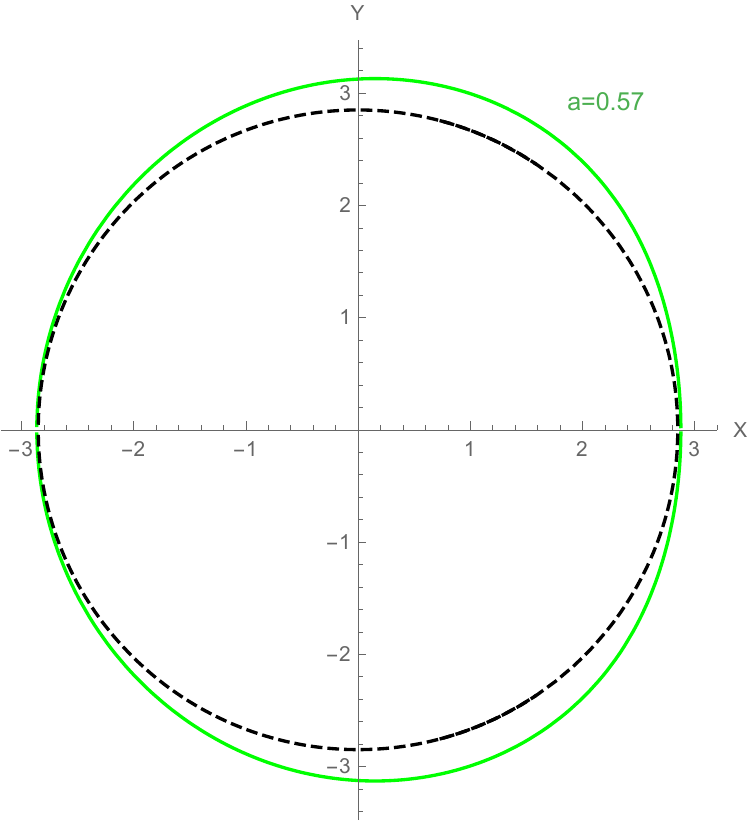}
       \includegraphics[width=2.8in]{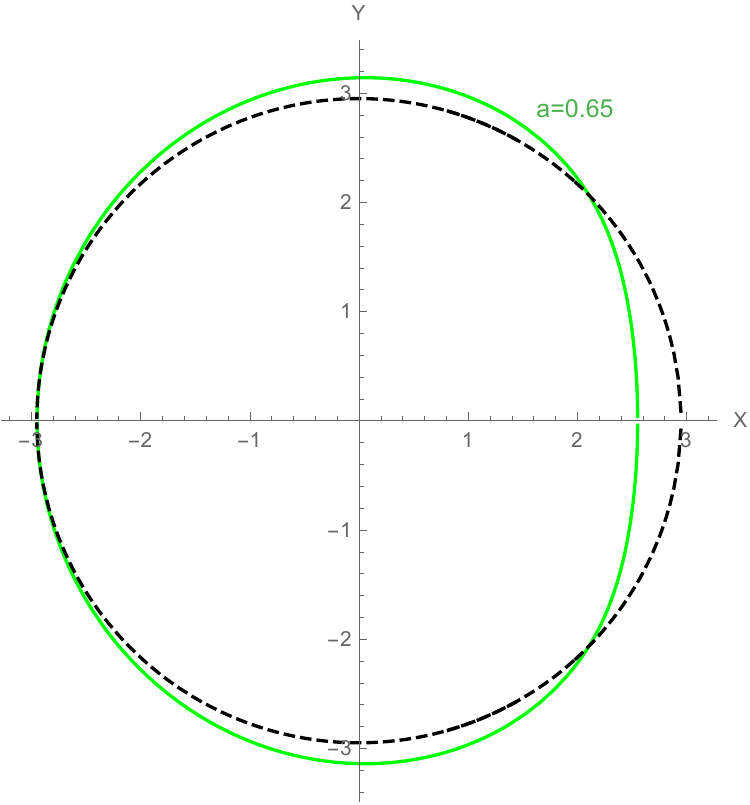}
        \caption{Hypershadow corresponding to a $Z=0$ ($v=0$) slice with $\theta_0=\pi/2,w_q=-0.3,c=0.1,M=2$. Top: left $a = 0.1$, right $a = 0.5$.
Bottom: left $a = 0.57$, right $a = 0.65$.}\label{FigAchatada}
  \end{center}
\end{figure}

It is also worth investigating the shape that the shadow takes as the rotation becomes stronger. In Figure \ref{FigAchatada}, we observe that for a weak rotation parameter ($a = 0.1$ in the example), the shadow is almost a perfect circle. As the rotation increases (in the example, for $a = 0.5$ and $a = 0.57$), the distance from the center of the shadow to its highest point along the vertical axis deviates from the radius of the circle, and thus the shadow no longer has a circular shape. As the rotation continues to increase, the shadow becomes a closed curve, as shown in the figure at the bottom right ($a = 0.65$ in the example). It is worth mentioning that, under the method we are using, for a positive spin value $a = 0.65$, the bottom right panel of Fig. 10 shows the dent or shadow distortion on the right side, contrary to what occurs in the usual Kerr shadow (see, for example, References \cite{Papnoi:2014aaa,Perlick:2021aok}).

In this way, we can observe that, under our methodology and given the influence of quintessence parameters in $5D$, the shape of our shadow differs from the $4D$ quintessence case \cite{Singh:2017xle}, as well as from other models studied in $5D$ \cite{Papnoi:2014aaa,Ahmed:2020jic,Amir:2017slq}. In the latter cases, the shape of the shadow is generally such that its right half resembles a circle, while the left half takes the form of a closed curve. Therefore, the study of observables such as the shadow radius and the distortion parameter should require a more detailed analysis. Despite this, we will take as a reference the definitions of these observables as given in references \cite{Hioki:2009na,Singh:2022dqs}. The shadow radius is difined as:
\begin{equation}
    R_s = \frac{(\bar{X}_r - \bar{X}_t)^2 + Y_t^2}{2|\bar{X}_r - \bar{X}_t|}
\end{equation}
where $(\bar{X}_t, Y_t)$ and $(\bar{X}_r, Y_r)$ represent the topmost and rightmost positions, respectively, of the celestial coordinates through which the contour of the black hole shadow passes.

Given that, in Figures \ref{FigSombras1} and \ref{FigAchatada}, the left and right radial horizontal distances coincide with respect to the vertical axis located at $\bar{X} = 0$, and that the farthest point from the center $(0,0)$ corresponds to $(0, Y_t)$, we estimate the distortion parameter as follows:
\begin{equation}
    \delta= \frac{|Y_t-\bar{X}_r|}{R_s}
\end{equation}

Furthermore, we provide a speculative methodology to test the shadow sizes in five dimensional scenarios, in light of the constraints provided by the Event Horizon Telescope (EHT) regarding the shadow size of the four-dimensional supermassive black hole M87. These constraints were measured at the $1\sigma$ (68\%) confidence level. The shadow size of this black hole is found to lie within the following interval \cite{Nozari:2024jiz,Cao:2023ppv,EventHorizonTelescope:2021dqv}:
\begin{equation} \label{Restriccion4D}
    4.31 \leq \frac{2}{V_{\text{vac}}(r = R_{\text{s}})} = \frac{R_{\text{s}}}{M} \leq 6.08
\end{equation}
where $V_{\text{vac}}(r)$ denotes the absolute value of the gravitational potential in the static scenario:
\begin{equation}
   V_{\text{vac}}(r) = \frac{2M}{r} 
\end{equation}
Here, strictly speaking, $M = G \cdot m$, with $G$ being Newton’s gravitational constant, which in Planck units has dimensions $[G] = \ell^2$, and $[m] = \ell^{-1}$. Consequently, the ratio $r/M$ is dimensionless.

In our case, since we are considering a five-dimensional scenario, we have $M = G_{5D} \cdot m$, where $[G_{5D}] = \ell^3$. The vacuum potential is given by:
\begin{equation}
   V^{(5D)}_{\text{vac}}(r) =  \frac{2M}{r^{2}}
\end{equation}
Thus, in a speculative way, we rewrite the constraint in Eq.~\eqref{Restriccion4D} as:
\begin{equation} \label{RestriccionLL}
    4.31 \leq \frac{2}{ V^{(5D)}_{\text{vac}}(r = R_{\text{s}})} = \frac{r_{\text{sh}}^{2}}{M}=\mbox{Ratio}  \leq 6.08
\end{equation}
It is straightforward to verify that the ratio $\left( \frac{r_{\text{sh}}^{2}}{M} \right)$ is dimensionless. 

For simplicity, we take the data shown in Figure \ref{FigSombras1} as a reference. In Table \ref{Tabla1}, we have computed the values of the shadow radius and the distortion parameter using the parameters from the aforementioned figure. We observe that, for fixed values of $w_q$, $a$, and $M$, the ratio in condition \eqref{RestriccionLL} can satisfy the constraints of the same equation, depending on the value of the quintessence parameter $c$. In the example of Table \ref{Tabla1} this occurs for $c \in \sim [0.223,0.4]$. This suggests that, depending on the strength of the quintessence term, the values of the static gravitational potential in $5D$ can be such that the theoretical results remain consistent with the experimental measurements for M87 obtained by the EHT in $4D$ scenarios without quintessence.

\begin{table}[h!]
\centering
\begin{tabular}{|c|c|c|c|c|c|c|c|}
\hline
$c$ & 0.12 & 0.223 & 0.285 & 0.335 & 0.4 & 0.95 & 2 \\
\hline
$\bar{X}_r$ & 5.2 & 4.8 & 4.55 & 4.35 & 4.1 & 2.75 & 1.75 \\
\hline
$\bar{X}_t$ & 0 & 0 & 0 & 0 & 0 & 0 & 0 \\
\hline
$Y_t$ & 5.45 & 4.9 & 4.65 & 4.5 &4.25  & 2.85 & 1.75 \\
\hline
$R_s$ & 5.456 & 4.901 & 4.651 & 4.503 & 4.200 & 2.852 & 1.750 \\
\hline
$\delta$ & 0.046 & 0.020 & 0.022 & 0.033 & 0.036 & 0.035 & 0.000 \\
\hline
Ratio & 7.442 & 6.005 & 5.408 & 5.068 & 4.410 & 2.033 & 0.766 \\
\hline
\end{tabular}
\caption{Values of $R_s$, $\delta$ and Ratio for $\theta_0=\pi/2,a=0.49,w_q=-0.3,M=4$.}
\label{Tabla1}
\end{table}

\subsection{Constraints from EHT observations of M87 using the circularity deviation criterion}

The Event Horizon Telescope (EHT) collaboration published the first image of the supermassive black hole in M87 in reference \cite{EventHorizonTelescope:2019ths}, enabling the investigation of possible constraints on our spacetime geometry based on the measurements of the black hole shadow provided by this collaboration. Specifically, these measurements revealed an asymmetric bright emission ring with a deviation from circularity of $\Delta C \leq 0.1$. The analysis of the shadow radius constraints presented above, based on the 5D Newtonian potential, provides a theoretical perspective. However, a more robust measure, since it does not depend on a dimensional analysis like the previous one, is the circularity deviation.

The contour of the shadow can be represented using polar coordinates $\left (R(\Phi), \Phi \right )$, taking as a reference the point $(\bar{X}_c, Y_c)$ such that $\bar{X}_c = \dfrac{\bar{X}_r - \bar{X}_l}{2}$ and $Y_c = 0$. In the last expression, $\bar{X}_l$ corresponds to  the leftmost point where the shadow contour and the reference circle meet the horizontal axis. On the other hand, the angle $\Phi$ is defined as: \begin{equation}
    \Phi \equiv \arctan \left ( \frac{Y}{\bar{X}-\bar{X}_c} \right )
\end{equation} Thus, $\Phi$ denotes the angle defined between the horizontal axis and the vector connecting the shadow center $(\bar{X}_c, Y_c)$ with the point $(\bar{X}, Y)$ located on the contour circumference. We follow the methodology proposed in Reference \cite{Bambi:2019tjh} (see also References \cite{Afrin:2021imp,Afrin:2021wlj,Ahmed:2025zdc}). In that work, the average shadow radius is defined as: \begin{equation} \label{RadioMedio}
    \bar{R}^2=\frac{1}{2\pi} \int_0^{2\pi} R^2(\Phi) d\Phi
\end{equation}where \begin{equation}
    R(\Phi) = \sqrt{(\bar{X}-\bar{X}_c)^2+(Y-Y_c)^2}
\end{equation} Reference \cite{Bambi:2019tjh} defines the circularity deviation $\Delta C$, which estimates the deviation from a perfect circular shape, as: \begin{equation} \label{DesviacionCircular}
    \Delta C =\frac{1}{\sqrt{2\pi} \cdot \bar{R}} \sqrt{\int_0^{2\pi} \left ( R(\Phi)- \bar{R}  \right)^2 d \Phi}
\end{equation}  In equations \eqref{PlotX} and \eqref{PlotY}, we can observe that $\bar{X} = \bar{X}(x)$ and $\bar{Y} = \bar{Y}(x)$. Numerically, we can determine the values of the coordinate $x$ such that $\bar{Y}(x_l) = \bar{Y}(x_r) = 0$. The resulting values depend on the specific values of the parameters involved. This allows us to obtain $\bar{X}_l = \bar{X}(x_l)$ and $\bar{X}_r = \bar{X}(x_r)$, and consequently $\bar{X}_c$. With this data, we compute the average shadow radius and the deviation from a perfect circular shape, equations \eqref{RadioMedio} and \eqref{DesviacionCircular}, respectively. For simplicity, we take the data shown in Figure \ref{FigSombras} as a reference. Additionally, we will include in our analysis the cases where $c = 0.05$ and $c = 0$, in order to test the limit of $\Delta C$ for $c \to 0$. As observed in Table \ref{TablaValores2},  for the quintessence parameter domain $c \in [0, 2]$, and keeping the remaining parameters fixed, the circular deviation $\Delta C$ decreases within the range $[0.0162, 0.0045]$. From this behavior, we can extrapolate that, for the chosen parameter set, $\Delta C \leq 0.1$, which is consistent with the EHT M87 constraints. This analysis suggests that, by employing the method for representing shadows in $5D$ as described in Reference \cite{Novo:2024wyn}, the results satisfy the bound $\Delta C \leq 0.1$ for the case with quintessence, as well as in the limiting case without quintessence.

\begin{table}[h!]
\centering
\begin{tabular}{|c|c|c|c|c|c|c|c|c|c|}
\hline
$c$ & 0 & 0.05 & 0.12 & 0.223 & 0.285 & 0.335 & 0.4 & 0.95 & 2 \\
\hline
$\bar{X}_r$ & 4.755 & 4.599 & 4.389 & 4.1 & 3.937 & 3.811 & 3.654 & 2.629 & 1.681 \\
\hline
$\bar{X}_l$ & -6.843 & -6.479 & -6.009 & -5.398 & -5.071 & -4.829 & -4.540 & -2.952 & -1.787 \\
\hline
$\bar{X}_c$ & -1.044 & -0.940 & -0.810 & -0.649 & -0.567 & -0.509 & -0.443 & -0.162 & -0.053 \\
\hline
$\bar{R}$ & 5.935 & 5.666 & 5.314 & 4.846 & 4.591 & 4.399 & 4.168 & 2.818 & 1.745 \\
\hline
$\Delta C$ & 0.0162 & 0.0159 & 0.0153 & 0.0142 & 0.0134 & 0.0128 & 0.0120 & 0.0069 & 0.0045 \\
\hline
\end{tabular}
\caption{Values of $\bar{X}_r$, $\bar{X}_l$, $\bar{X}_c$, $\bar{R}$ y $\Delta C$ for $\theta_0=\pi/2,a=0.49,w_q=-0.3,M=4$.}
\label{TablaValores2}
\end{table}
%
%
%
%
%
%%%%%%%%%%%%%%%%%%%%%%% ENERGY CONDITIONS %%%%%%%%%%%%%%%%%%%%%
\section{Energy conditions}

In this section, the energy conditions for the rotating and spherically symmetric black hole in five dimensions are analyzed. We establish the procedure to be used to determine the components of the energy density and effective pressures. Based on this, we will analyze the values of the geometry's parameters in which the energy conditions are satisfied or violated. It is worth noting that we will use the following definitions to analyze the energy conditions:
\begin{align}
\text{Weak Energy Condition (WEC): }& ~\rho_q \geqslant 0,~\rho_q + p_i \geqslant 0
\label{ec1}\\
\text{Null Energy Condition (NEC): }& ~\rho_q + p_i \geqslant 0  \label{ec2}\\
\text{Strong Energy Condition (SEC): }& \rho_q + p_i \geqslant 0, ~ \rho_q +\sum p_i\geqslant 0
\label{ec3}\\
\text{Dominant Energy Condition (DEC):} & 
~\rho_q \geqslant 0, \rho_q - |p_i|\geqslant  0 \label{ec4}
\end{align}
For our line element, the structure of the energy-momentum tensor components is described in detail in Appendix \ref{ComponentesEM} for the mass function \eqref{FuncionDeMasa}. On the one hand, it is well known that the Myers-Perry solution, which corresponds to a $5D$ generalization of the Kerr solution, is vacuum in nature, so it makes no sense to study the energy conditions. In the rotating $5D$ case with presence of matter sources in the energy momentum tensor, due to the non-diagonal structure of the energy-momentum tensor, it becomes difficult to study the energy conditions. An example of how to address this problem can be seen in reference \cite{Amir:2020fpa}, where a one-form of the dual basis is defined, such that a diagonal energy-momentum tensor is obtained at the locations $\theta = 0$. In this work, similarly to the last reference, we will use the basis from reference \cite{Aliev:2004ec} to obtain a diagonalized version of the energy-momentum tensor. This will allow us to analyze the energy conditions.

As mentioned above, to obtain the components of the energy-momentum tensor associated with energy density and pressures we consider a locally nonrotating observer moving in the $\phi$-plane. For this purpose, we can use the following basis \cite{Aliev:2004ec}
\begin{align}
%    e^{(t)} &= \sqrt{g_{tt} - \frac{g_{t\psi}^2}{g_{\psi\psi}} - \Omega^2 \frac{g_{\phi\phi}g_{\psi\psi}-g_{\phi\psi}^2}{g_{\psi\psi}}}dt,\nonumber\\
    e^{(t)} &= \sqrt{\bigg |g_{tt} - \frac{g_{t\psi}^2}{g_{\psi\psi}} - \Omega^2 \frac{g_{\phi\phi}g_{\psi\psi}-g_{\phi\psi}^2}{g_{\psi\psi} }\bigg |}dt=\sqrt{\pm \bigg (g_{tt} - \frac{g_{t\psi}^2}{g_{\psi\psi}} - \Omega^2 \frac{g_{\phi\phi}g_{\psi\psi}-g_{\phi\psi}^2}{g_{\psi\psi} }\bigg )}dt,\nonumber\\
    e^{(r)} &= \sqrt{g_{rr}}dr, \nonumber \\
    e^{(\theta)} &=\sqrt{g_{\theta\theta}}d\theta, \nonumber \\
    e^{(\phi)} &= - \sqrt{\frac{g_{\phi\phi}g_{\psi\psi}-g_{\phi\psi}^2}{g_{\psi\psi}}}(\Omega dt - d\phi),\nonumber \\
    e^{(\psi)} &= \sqrt{g_{\psi\psi}}\left( \frac{g_{t\psi}}{g_{\psi\psi}}dt + \frac{g_{\phi\psi}}{g_{\psi\psi}}d\phi + d\psi\right),
\end{align}

Where the signs $+$ or $-$ of the component $e^{(t)}$ correspond to when the expression inside the parentheses is positive or negative, respectively. In the case of an observer located after the event horizon, the minus sign should be chosen. This is clearly observed in the static case, where $g_{tt} < 0$ at this location.

The term $\Omega$ is given by
\begin{equation}
    \Omega = \frac{g_{t \psi}g_{\phi\psi}-g_{t\phi}g_{\psi \psi}}{g_{\phi\phi}g_{\psi\psi} - (g_{\psi\psi})^2}.
\end{equation}
As a consequence, the effective energy-momentum tensor $T^{\mu\nu}$ can be seen as a projection associated with the chosen tetrad basis in the form
\begin{equation}
    T^{(a)(b)} = e_{\mu}^{(a)} e_{\nu}^{(b)} T^{\mu\nu}.
\end{equation}
Next, we can use the connection between $T^{\mu\nu}$ and the Einstein tensor, such that the energy-momentum tensor  $T^{(a)(b)}$ becomes 
\begin{equation}
    8\pi T^{(a)(b)} = e_{\mu}^{(a)} e_{\nu}^{(b)} G^{\mu\nu}.
\end{equation}
The resulting expressions are rather lengthy; hence, setting $\theta = 0$ allows us to obtain more compact expressions, which can be written as follows:
\begin{align}  
8\pi T^{(t)(t)} &=  - \frac{{\Delta} ~r~b^2\frac{d^2}{dr^2} m(r)  -  \left(\frac{d}{dr} m(r)\right)  (b^2 + r^2)^2 (a^2 + 3r^2)}
{2 (a^2 + r^2) r^3 [(a^2+r^2)(b^2+r^2) + b^2~m(r)]} , \label{Ttt}\\
    8\pi T^{(r)(r)} &= - \frac{\left( \frac{d}{dr} m(r) \right) (a^2 + 3r^2)}
{2r (a^2 + r^2)^2}, \label{Trr} \\
    8\pi T^{\:\:(\theta)(\theta)} &= - \frac{\frac{r (a^2 + r^2)}{2}  \frac{d^2}{dr^2} m(r) + a^2  \frac{d}{dr} m(r) } 
{r (a^2 + r^2)^2}, \label{Ttheta} \\
    8\pi T^{(\phi)(\phi)} &=- \frac{\left(\frac{d^2}{dr^2}m(r)\right)a^2r + \left(\frac{d^2}{dr^2}m(r)\right)r^3 + 2a^2\frac{d}{dr}m(r)}{2r(a^2+r^2)^2}, \\        
    8\pi T^{(\psi)(\psi)} &=- \frac{(b^2 + r^2)^2 (a^2 + r^2)^2 r \frac{d^2}{dr^2} m(r) - b^2 \left(\frac{d}{dr} m(r)\right) (a^2 + 3 r^2){\Delta}}{2 (a^2 + r^2)^2 r^3 [(a^2+r^2)(b^2+r^2) + b^2~m(r)]},
\end{align}
{where, according to equation \eqref{Delta1}, $\Delta=(a^2+r^2)(b^2+r^2) - r^2~m(r)$}. While our analysis employs the specific mass function given by Eq.  \eqref{FuncionDeMasa}, these above equations are general conditions associated with an arbitrary mass function.

In a more specific analysis, for the case where a = b, we can also point out the following:

\begin{enumerate}
    \item Energy Density: In Figure \ref{Ttt_M}, we can see that for a fixed value of $w_q = -0.35$, $\rho_q \geq 0$ for the entire range of the radial coordinate. In Figure \ref{nec_violation}, we observe that for the parameters previously analyzed, with $w_q$ variable and the other parameters fixed, such that the geometry represents a rotating black hole ($w_q \geq -0.5$), $\rho_q \geq 0$ is always satisfied. For the parameters mentioned, with $w_q < -0.5$, for which the geometry represents a naked singularity, the WEC condition and DEC are violated near the radial origin. The aforementioned can also be observed in figures \ref{contour_Ttt_M}, \ref{rho_wq}.
\begin{figure}[ht]
    \begin{center}
      \includegraphics[width=3.5in]{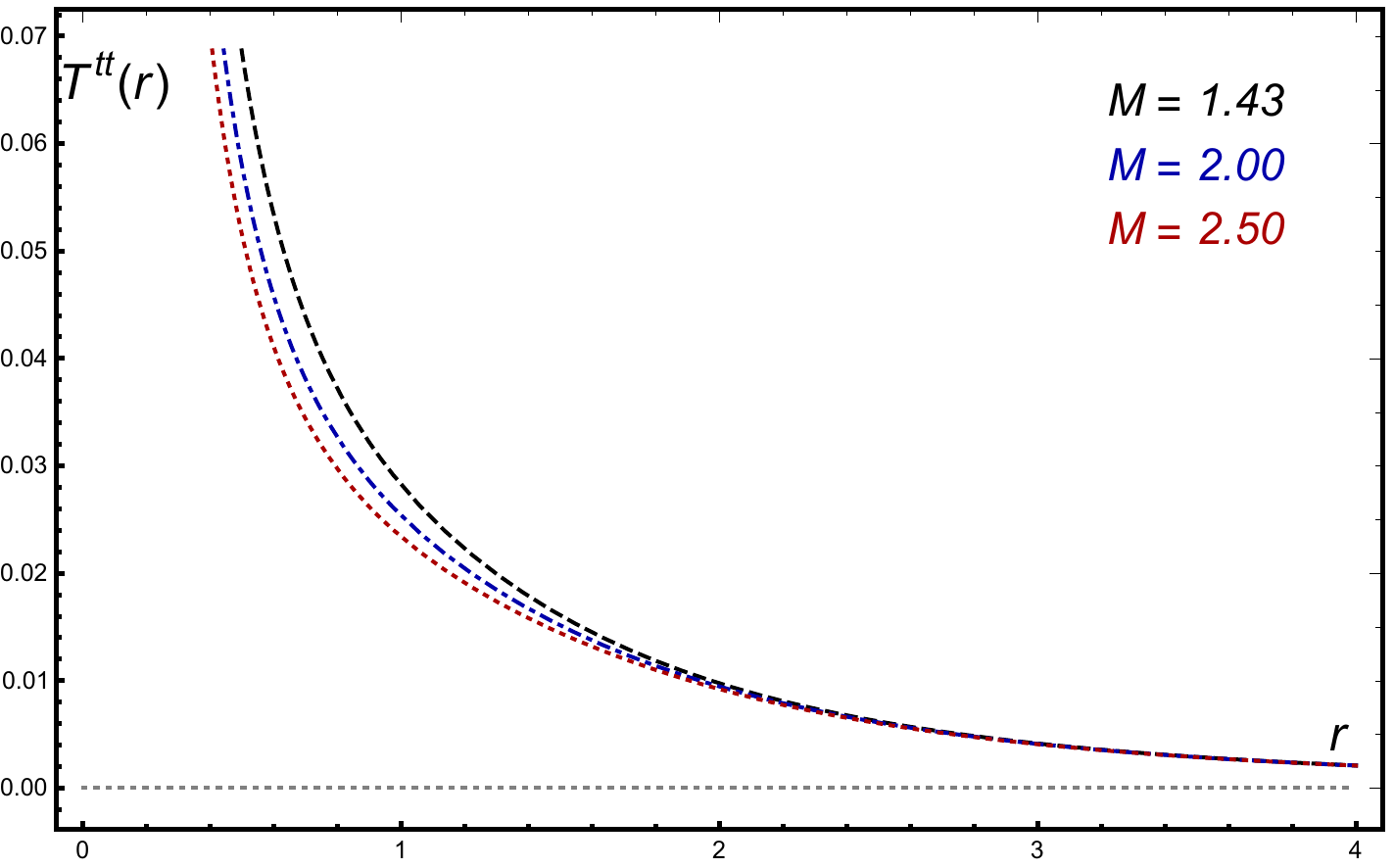}
    \caption{Plot illustrating $T^{(t)(t)}(r)$ [Eq. \eqref{Ttt}] for $M=1.43$ (dashed black line), $M=2.00$ (dash-dotted blue line), and $M=2.50$ (dotted red line). Here we adopted $a = b = c = 1$ and  $w_q = -0.35$. Since $\rho_q=T^{(t)(t)} \geqslant 0$, WEC conditions are satisfied in this particular case.}\label{Ttt_M}
    \end{center}
\end{figure}
\begin{figure}[ht]
  \begin{center}
      \includegraphics[width=3.8in]{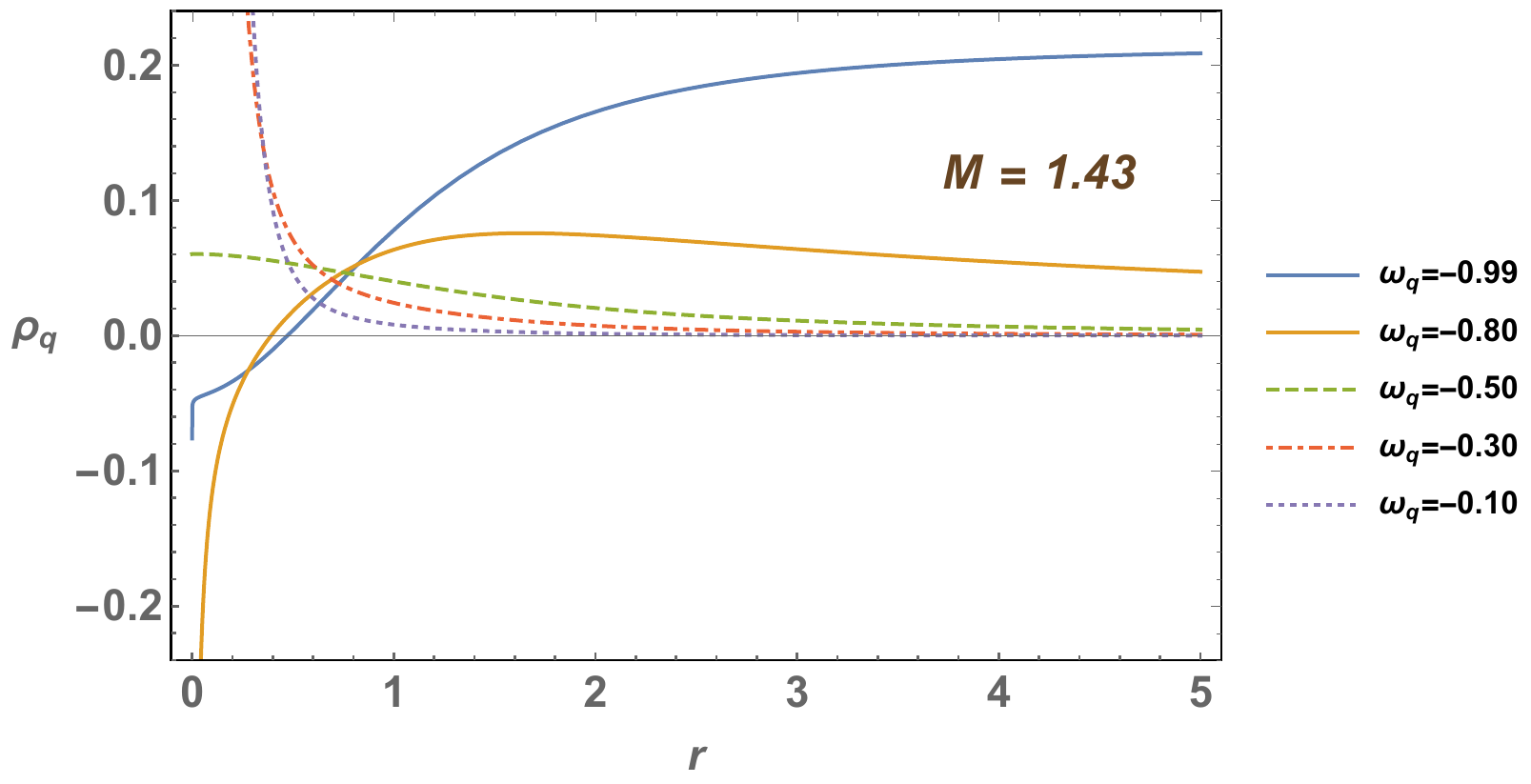}
      \includegraphics[width=3.06in]{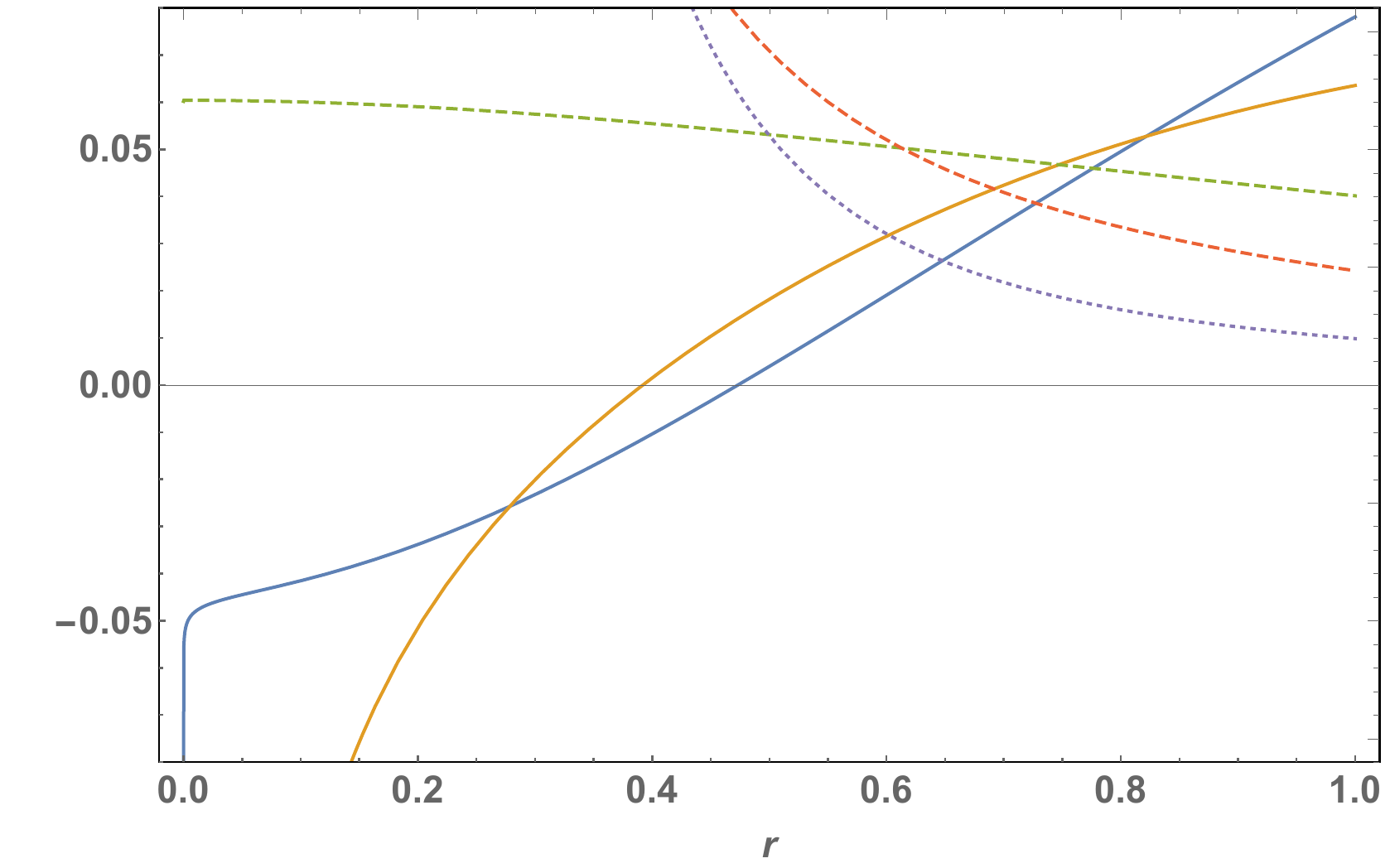}
  \caption{The left figure shows energy density $\rho_q(r)$ for several values of $w_q$ with $0 \leqslant r \leqslant 5$. On the right, a zoom into the small region near the origin ($0 \leqslant r \leqslant 1.0$). Here we use $a = b = c = 1$, and  $M = 1.43$. WEC conditions are violated for $w_q<-0.5$, as evidenced by negative energy densities in this regime}.\label{nec_violation}
  \end{center}
\end{figure}

\begin{figure}[ht]
  \begin{center}
    \includegraphics[width=4in]{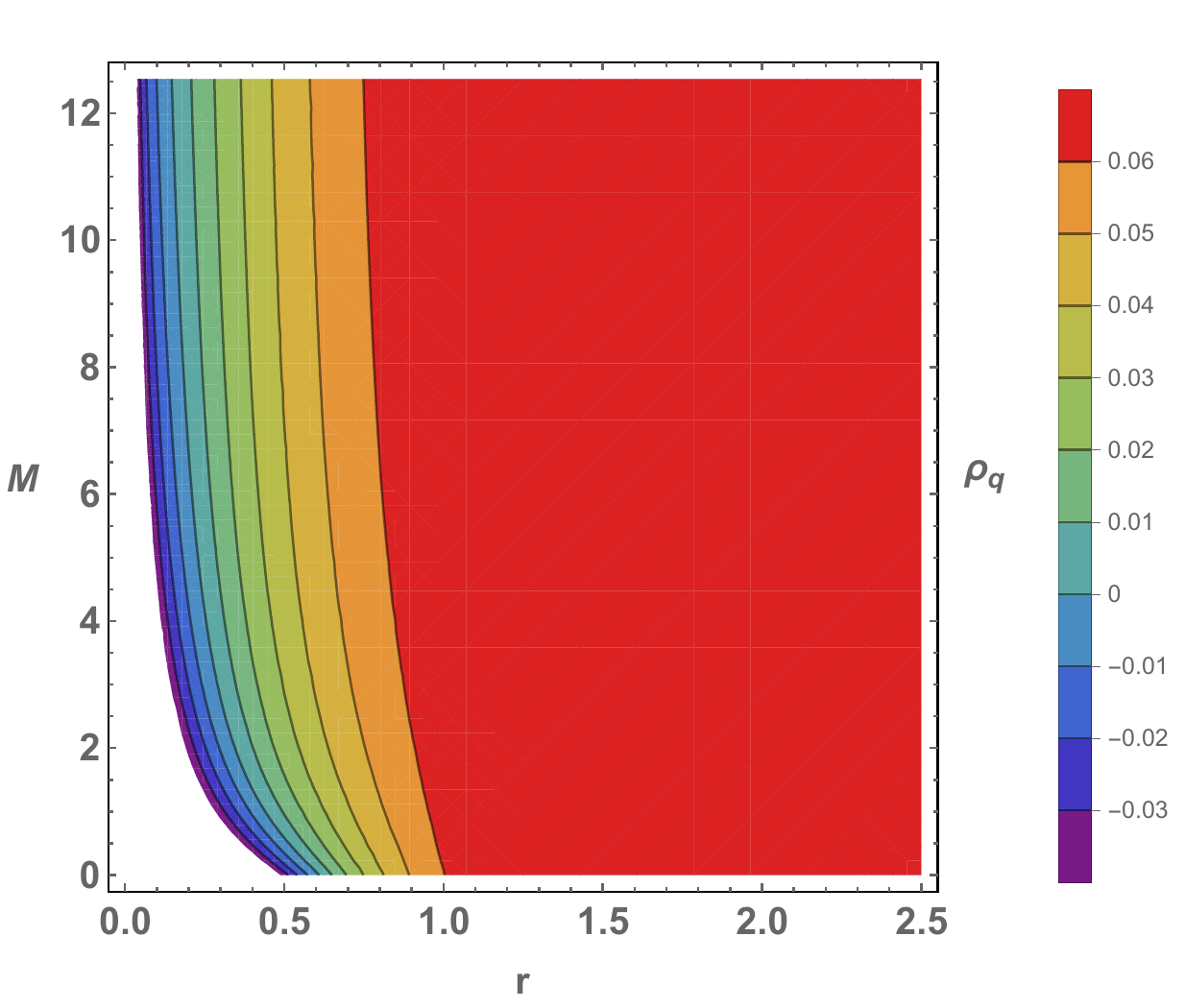}
  \caption{ Contour plot illustrating $\rho_q=T^{(t)(t)}$ [see Eq. \eqref{Ttt}], varying $M$ (vertical left axis) for $r \geqslant 0$. Near the origin, WEC are violated whenever $w_q \leq -0.50$.  Here we adopted $w_q = -0.80$, $a = 1$, $b = 1$,  and $c = 1$.}\label{contour_Ttt_M}
  \end{center}
\end{figure}

\begin{figure}[ht]
  \begin{center}
      \includegraphics[width=4in]{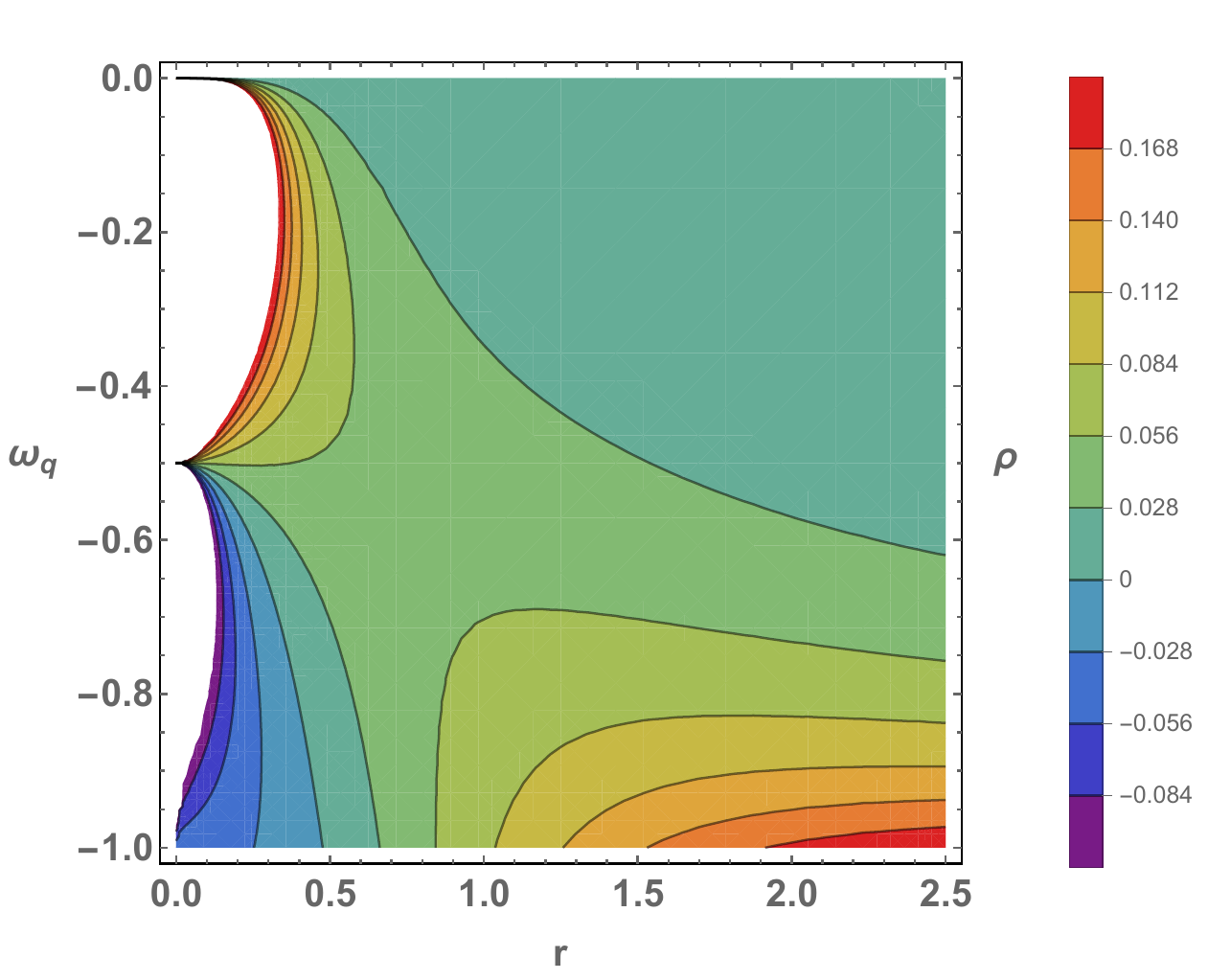}
  \caption{Contour plots illustrating $\rho_q(r)$ (vertical right axis)  varying $w_q$ (vertical left axis) for $r \geq 0$ (horizontal axis). Here we adopted $a = 1$, $b = 1$,  $c = 1$, and  $M = 1.43$. Near the origin, WEC conditions are violated ($\rho_q<0$) { whenever} $w_q<-0.5$.}\label{rho_wq}
  \end{center}
\end{figure}
\item WEC and NEC: From the previous item, we have that $\rho_q \geq 0$ is satisfied for the parameters where the geometry represents a rotating black hole. Therefore, to test regions where the WEC is satisfied, we still need to test the regions where $\rho_q + p_i \geqslant 0$, i.e., where NEC is satisfied. To test the latter, from now on, we will focus on the parameters where the geometry represents a rotating black hole.
\begin{itemize}
    \item $\rho_q+p_r$: In Figure \ref{FigRhoPrM}, we observe the behavior of $\rho_q + p_r$ for different values of the parameter $M$ while the other parameters are fixed. In Figure \ref{FigRhoPrW}, we observe the behavior of $\rho_q + p_r$ for different values of the parameter $w_q$ with the other parameters fixed. In both cases, we observe a generic behavior moving from left to right in the radial coordinate: $\rho_q + p_r \geq 0$, $\rho_q + p_r \leq 0$, and $\rho_q + p_r \geq 0$, respectively. That is, there exists an intermediate zone in the radial coordinate where the NEC and WEC are violated.
\begin{figure}[hb]
  \begin{center}
      \includegraphics[width=3.in]{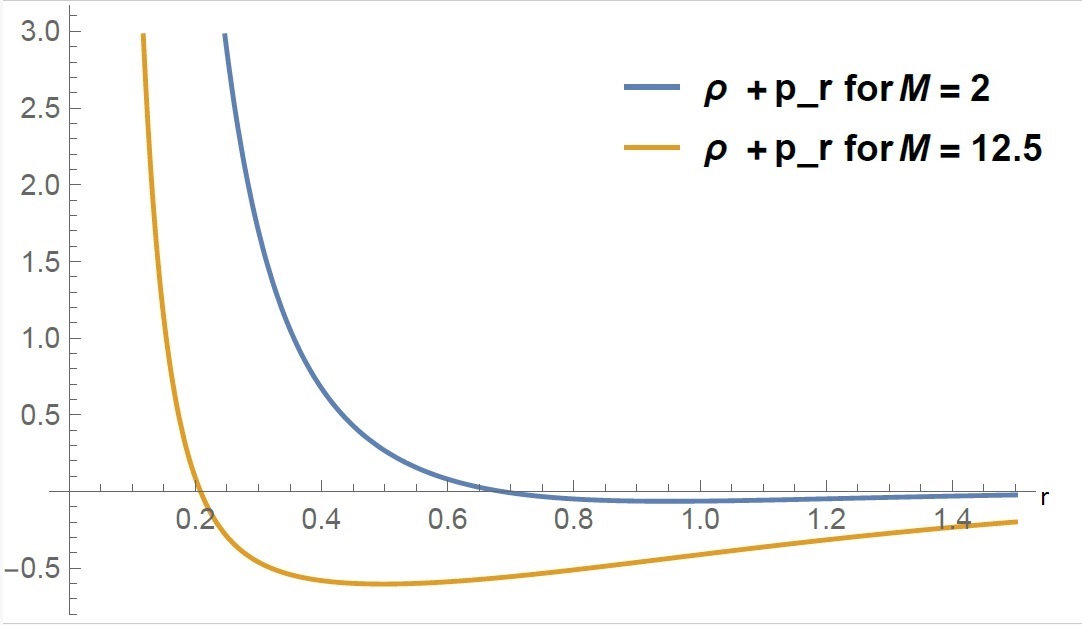}
      \includegraphics[width=3.in]{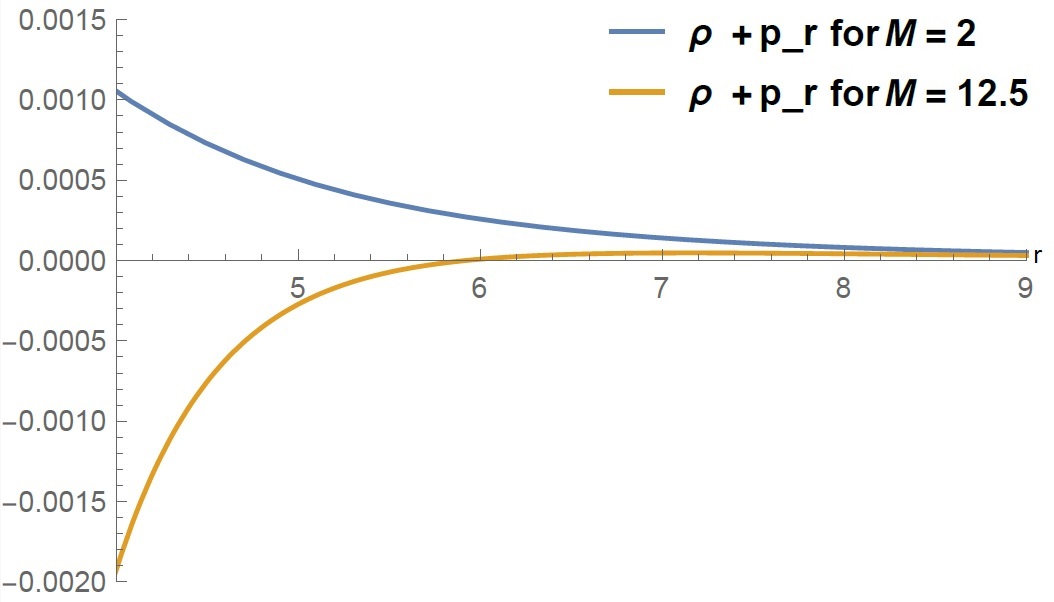}
  \caption{$\rho_q+p_r$ for $M=2$ and $M=12.5$ and $a=b=c=1,w_q=-0.35$}.\label{FigRhoPrM}
  \end{center}
\end{figure}

\begin{figure}[ht]
  \begin{center}
      \includegraphics[width=3.in]{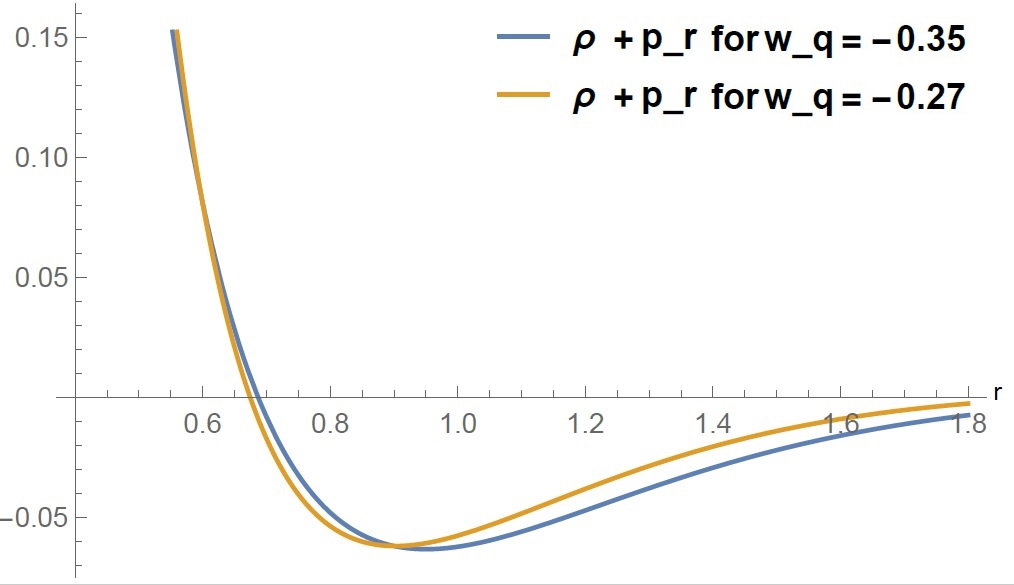}
      \includegraphics[width=3.in]{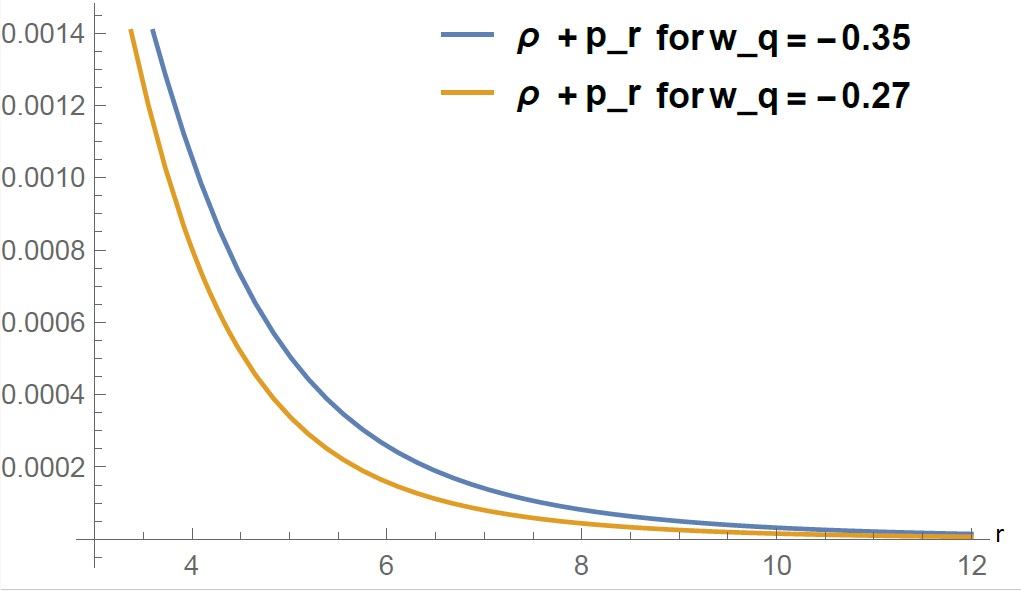}
  \caption{$\rho_q+p_r$ for $w_q=-0.27$ and $w_q=-0.35$ and $a=b=c=1,M=2$}.\label{FigRhoPrW}
  \end{center}
\end{figure}
\item  $\rho_q+p_\theta=\rho_q+p_\phi$: In Figure \ref{FigRhoPthetaM}, we observe the behavior of $\rho_q + p_\theta$ for different values of the parameter $M$ while the other parameters are fixed. In Figure \ref{FigRhoPthetaW}, we observe the behavior of $\rho_q + p_\theta$ for different values of the parameter $w_q$ with the other parameters fixed. In the first case, we observe a generic behavior moving from left to right in the radial coordinate: $\rho_q + p_\theta \geq 0$, $\rho_q + p_\theta \leq 0$, and $\rho_q + p_\theta \geq 0$, respectively. That is, there exists an intermediate zone in the radial coordinate where the NEC and WEC are violated. In the second case, we find that for the particular case with $w_q = -0.27$ and the other parameters mentioned in the figure, this condition is satisfied. However, for these latter parameters, it was tested in the previous item that there exists a region where $\rho_q + p_r \leq 0$, meaning that both WEC and NEC would also be violated in this region.

\begin{figure}[ht]
  \begin{center}
      \includegraphics[width=3.in]{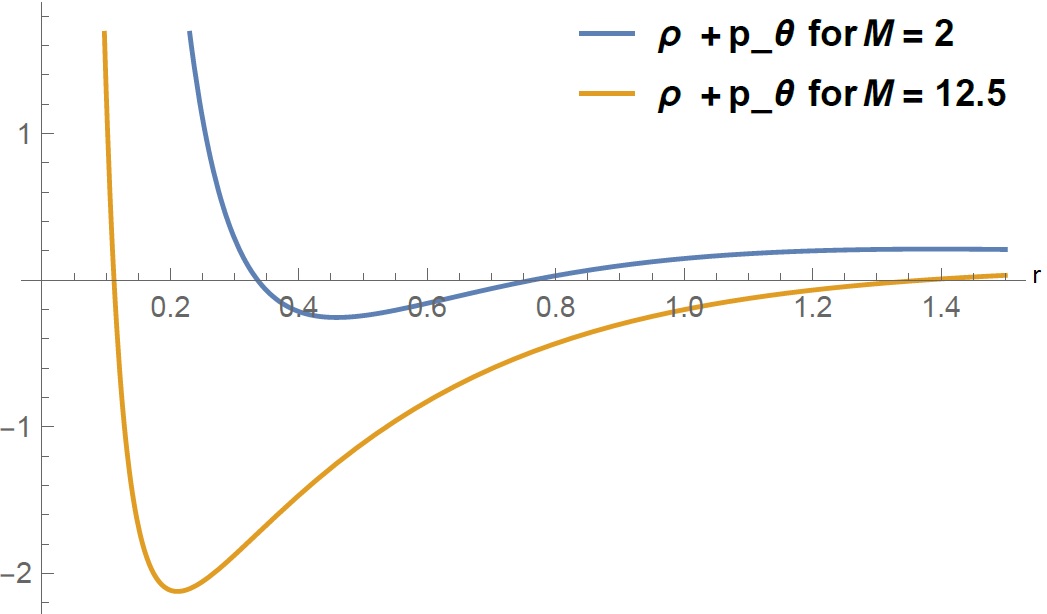}
      \includegraphics[width=3.in]{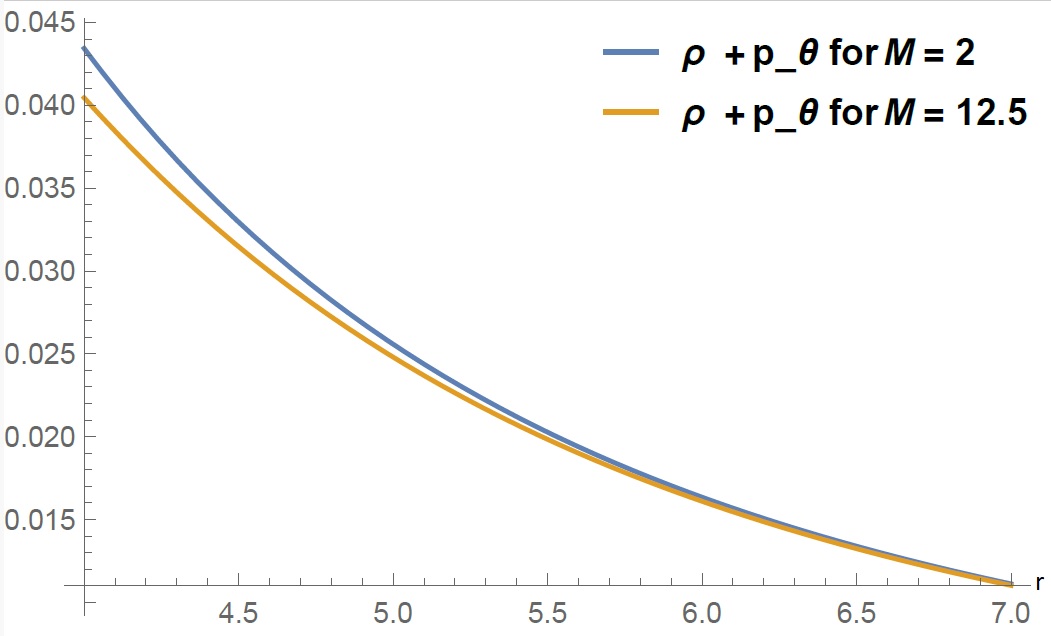}
  \caption{$\rho_q+p_\theta=\rho_q+p_\phi$ for $M=2$ and $M=12.5$ and $a=b=c=1,w_q=-0.35$}.\label{FigRhoPthetaM}
  \end{center}
\end{figure}

\begin{figure}[ht]
  \begin{center}
      \includegraphics[width=3.in]{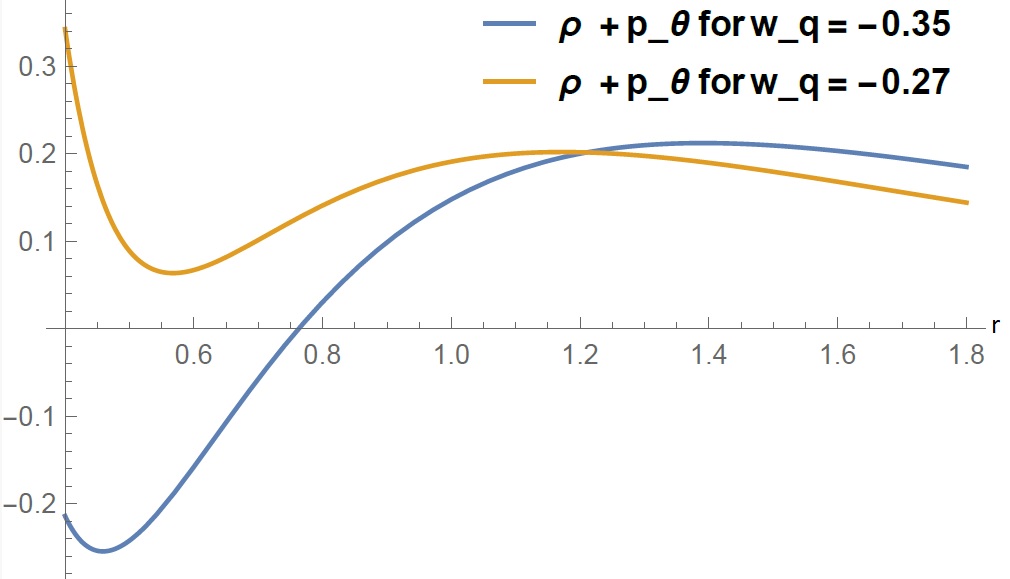}
      \includegraphics[width=3.in]{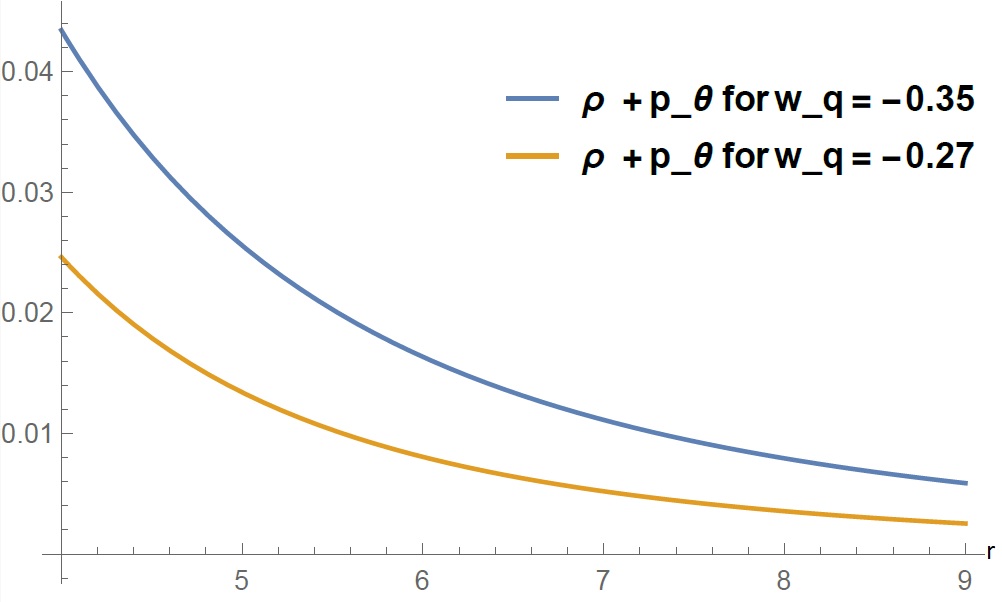}
  \caption{$\rho_q+p_\theta$ for $w_q=-0.27$ and $w_q=-0.35$ and $a=b=c=1,M=2$}.\label{FigRhoPthetaW}
  \end{center}
\end{figure}
\item  $\rho_q+p_\psi$: In Figure \ref{FigRhoPtsiM}, we observe the behavior of $\rho_q + p_\psi$ for different values of the parameter $M$ while the other parameters are fixed. In Figure \ref{FigRhoPtsiW}, we observe the behavior of $\rho_q + p_\psi$ for different values of the parameter $w_q$ with the other parameters fixed. In both figures, we observe that, with the chosen $w_q$ parameters, this condition is satisfied for values close to $M = 2$. However, since the other conditions from the previous items are violated in some ranges, the NEC and WEC are also violated in the mentioned ranges, which correspond to where the other conditions become negative.

\begin{figure}[ht]
  \begin{center}
      \includegraphics[width=3.in]{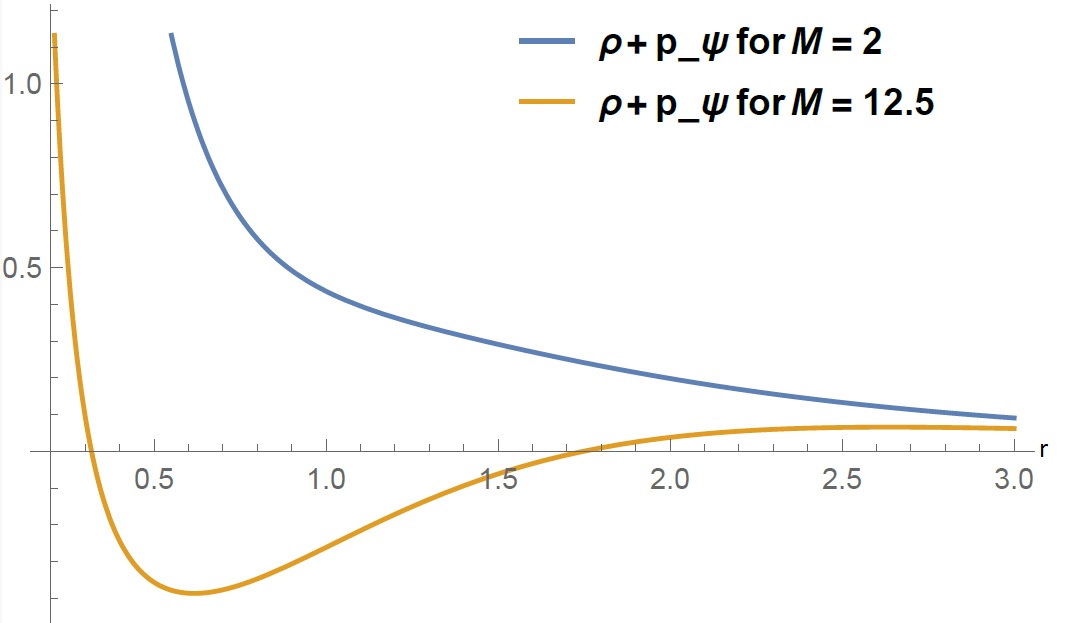}
      \includegraphics[width=3.in]{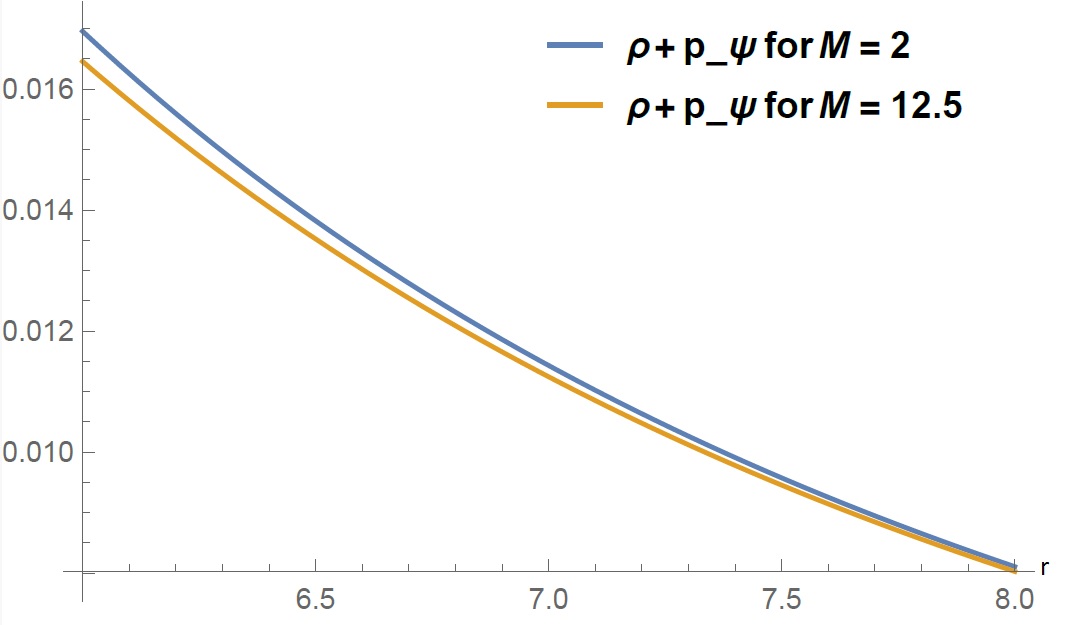}
  \caption{$\rho_q+p_\psi$ for $M=2$ and $M=12.5$ and $a=b=c=1,w_q=-0.35$}.\label{FigRhoPtsiM}
  \end{center}
\end{figure}

\begin{figure}[ht]
  \begin{center}
      \includegraphics[width=3.in]{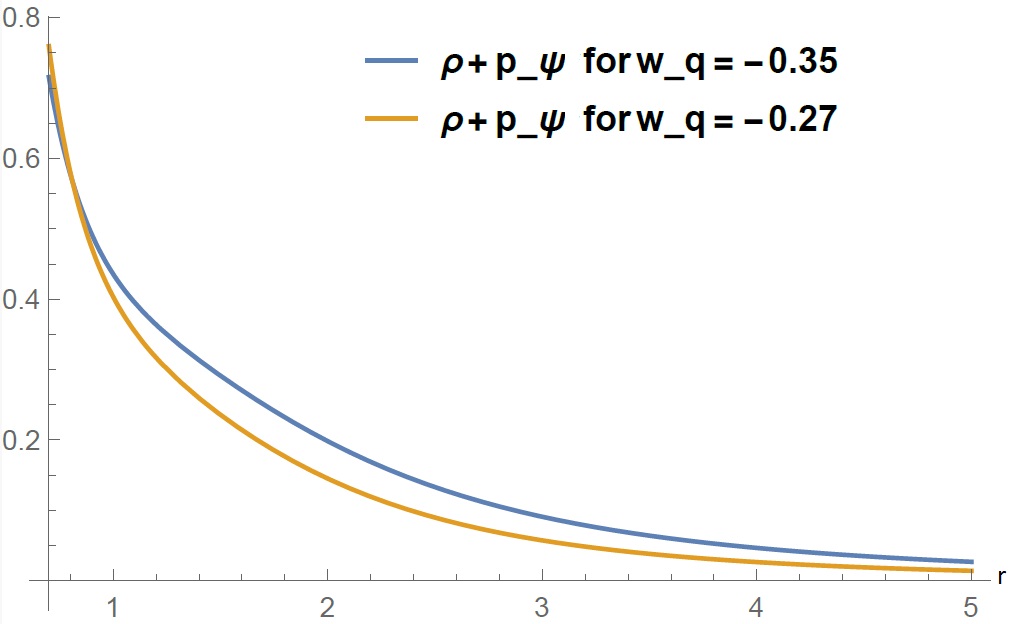}
        \caption{$\rho_q+p_\psi$ for $w_q=-0.27$ and $w_q=-0.35$ and $a=b=c=1,M=2$}.\label{FigRhoPtsiW}
  \end{center}
\end{figure}
\end{itemize}
In this way, we can state that for the chosen parameters, there are no cases where the energy conditions NEC and WEC are satisfied for all value of the radial coordinate. This is because it is sufficient for one condition to be negative in a specific range of the radial coordinate for the NEC and WEC to be violated in that range. It is also worth mentioning that, in the specific ranges where the NEC is violated, the SEC is also violated.

\item DEC: We will use values of $w_q$ such that the geometry represents a black hole, and as mentioned earlier, $\rho \geq 0$. In Fig \ref{FigDECM}, we observe the behavior of the DEC conditions by varying the parameter $M$ while keeping the other parameters fixed. We see that, for both $M=2$ and $M=12.5$, the radial segment where the DEC is violated is bounded by the intersection with the radial axis located furthest to the left in the second figure from the top (which corresponds to $\rho - |p_\theta|$) and by the intersection with the radial axis located furthest to the right in the first figure from the top (which corresponds to $\rho - |p_r|$). We also observe that this segment, where the DEC is violated, widens as the value of the parameter $M$ increases.

\begin{figure}[ht]
  \begin{center}
      \includegraphics[width=3.in]{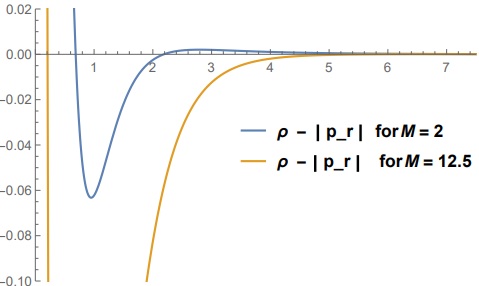}
      \includegraphics[width=3.in]{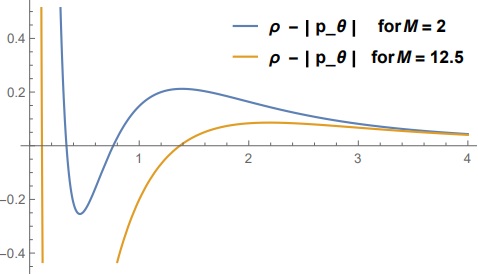}
      \includegraphics[width=3.in]{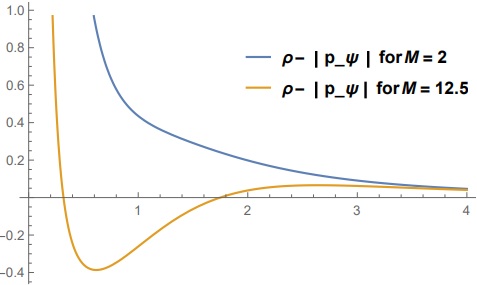}
  \caption{DEC for $M=2$ and $M=12.5$ and $a=b=c=1,w_q=-0.35$}.\label{FigDECM}
  \end{center}
\end{figure}

In Fig.\ref{FigDECW}, we observe the behavior of the DEC conditions by varying the parameter $w_q$, while keeping the other parameters fixed. We see that, for $w_q = -0.27$, the segment where the DEC is violated is determined by the first figure from the top (which corresponds to $\rho - |p_r|$). For $w_q = -0.35$, the segment where the DEC is violated is bounded by the intersection with the radial axis located furthest to the left in the second figure from the top (which corresponds to $\rho - |p_\theta|$) and the intersection with the radial axis located furthest to the right in the first figure from the top (which corresponds to $\rho - |p_r|$).

\begin{figure}[ht]
  \begin{center}
      \includegraphics[width=3.in]{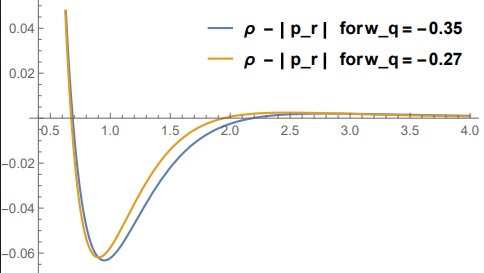}
      \includegraphics[width=3.in]{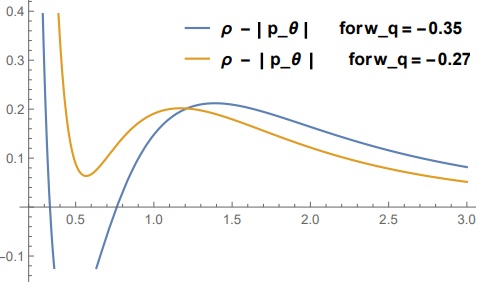}
      \includegraphics[width=3.in]{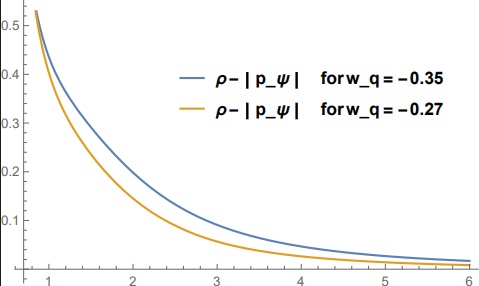}
  \caption{DEC for $w_q=-0.27$ and $w_q=-0.35$ and $a=b=c=1,M=2$}.\label{FigDECW}
  \end{center}
\end{figure}

\end{enumerate}

\begin{figure}[ht]
\includegraphics[scale=0.9]{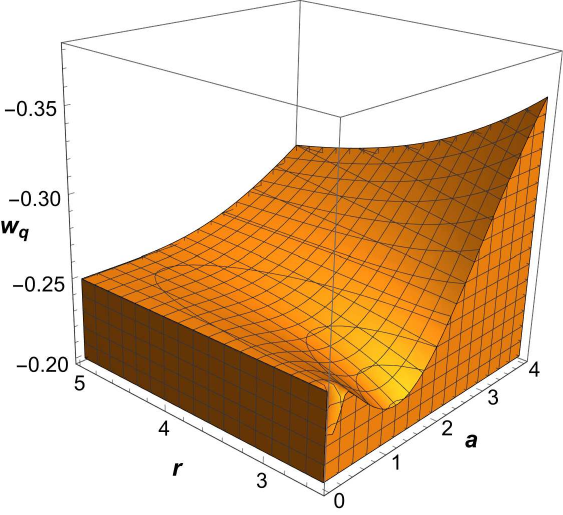}\newline
\caption{In this graph we plot the region in which the WEC, NEC, SEC, and DEC are satisfied. Where the parameter values are $b = a$, $c=1$, and $M=2$.}
\label{fig3d}
\end{figure}

\begin{figure}[ht]
\includegraphics[scale=0.9]{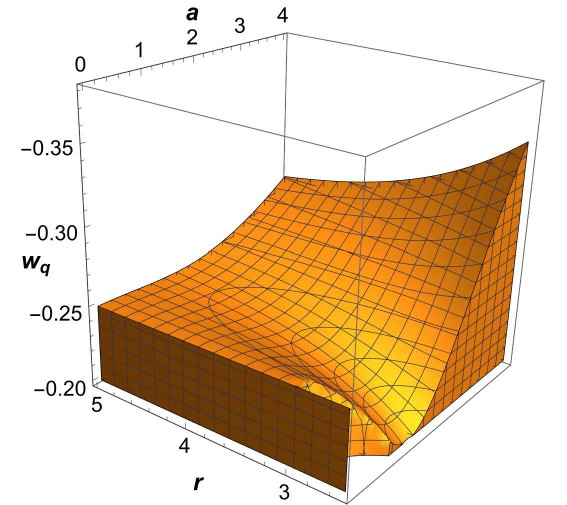}\newline
\caption{The figure plots the region in which the  WEC, NEC, SEC, and DEC are satisfied. Here, we consider the parameter values $b = a$, $c=1$, and $M=3$.}
\label{fig3d2}
\end{figure}

Despite the above, we will still search for values in the parameter space such that all energy conditions are satisfied. For this, we use the \texttt{RegionPlot3D} function from Mathematica to study these energy conditions. This function can plot a region associated with a combination of inequalities connected by logical operators. As we can see in Fig. \ref{fig3d}, all energy conditions (WEC,NEC,SEC and DEC) can be satisfied for the values $w_q \gtrsim -0.25$, all $a \geq 0$, and $r \gtrsim 5/2$. If $w_q \lesssim -0.25$, there is still a region where the energy conditions are satisfied corresponding to the values $a \gtrsim 2$ and $r \lesssim 5$. In the same way, figure \ref{fig3d2} has a similar behavior for the regions where the energy conditions are satisfied except in the region located at $\frac{5}{2} \lesssim r \lesssim \frac{7}{2}$, 
$0 \lesssim a \lesssim 2$, 
$-0.25 \lesssim w \lesssim -0.20$ where Figure \ref{fig3d2} has more irregular behavior than Figure \ref{fig3d}.

%
%\begin{figure}[h!]
%  \begin{center}
%      \includegraphics[width=4.in]{contour_plot_p_psi.pdf}
      %\includegraphics[width=4.in]{contour_plot_p_theta.pdf}
%  \caption{Contour plots illustrating the relationship between angular pressure $p_{\psi}(r)$ and  $\omega_q$ for $a = b = c = 1$ and  $M = 1.43$. \sout{WEC conditions are violated in this case for some range of $\omega_q$ values}}.
%  \end{center}
% \end{figure}

%\begin{figure}[h!]
%  \begin{center}
%      \includegraphics[width=4.in]{contour_plot_pr.pdf}
%      \includegraphics[width=4.in]{contour_plot_p_theta.pdf}
%  \caption{Contour plots illustrating the radial pressure $p_r(r)$ (top) and  the angular pressure $p_\theta$ (bottom) against $\omega_q$ for $a = b = c = 1$ and  $M = 1.43$. \sout{WEC conditions are violated in this case for some range of $\omega_q$ values}. {\color{red} [Eq \eqref{Trr} suggests that $p_r$ is mass-independent!]}} \label{contour_plot_pr}
%  \end{center}
% \end{figure}

\subsection{Regarding the case \texorpdfstring{$a \neq 0$, $b = 0$}{not}}

Although our work is focused on the case $a=b$ for the reasons mentioned earlier, it is worth discussing what happens in the case $b=0$. This is because in the latter case the WEC, NEC, and SEC can be analyzed analytically. In this case, the analytical study of the DEC turns out to be non--trivial and, therefore, will not be considered in this special case (however, it was considered numerically above). Remarkably, the analysis of the WEC, NEC, and SEC provides important information from an analytical perspective. . The conditions (\ref{ec1}), (\ref{ec2}), and (\ref{ec3}) reduce to
%\begin{align}
%    -cw_q &\geq 0,\\
%   -cw_q[w_q(a^2 + r^2) + r^2] &\geq 0,\\
%    -cw_q\left[w_q(a^2 + r^2) + \frac{a^2}{2} + r^2\right] &\geq 0,\\
%    -cw_q\left[w_q(a^2 + r^2) - \frac{a^2 - 3r^2}{12} \right] &\geq 0. 
% \end{align}
%
% Or more specifically:
%
\begin{align}
&  \rho_q \ge 0 \Rightarrow  -cw_q \geq 0,\\
&  \rho_q +p_r \ge 0 \Rightarrow  \,\, \mbox{automatically satisfied},\\
& \rho_q +p_\theta \ge 0 \Rightarrow    -cw_q\left [r^2(w_q + 1) + a^2\right ] \geq 0,\\
& \rho_q +p_\psi \ge 0 \Rightarrow    -cw_q\left[r^2 (1+w_q)+ a^2 (w_q+1/2) \right] \geq 0,\\
& \rho_q +p_r+ 2p_\theta +p_\psi {\geq 0} \Rightarrow   -cw_q\left[r^2 (w_q+1/4) + a^2 (w_q-1/12) \right] \geq 0. 
\end{align}
It is worth observing that in the particular case considered ($b=0$ and $\theta = 0$), the conditions $\rho_q \geq 0$, $\rho_q + p_r \geq 0$, $\rho_q + p_\theta = \rho_q + p_\phi \geq 0$ are trivially satisfied, given that $w_q \in [-1, 0]$. In the case of a physical singularity, which corresponds to $w_q \in [-1/2, 0]$, the condition $\rho_q + p_\phi \geq 0$ is also satisfied and, consequently, in this latter case, both the NEC and WEC are satisfied for all values of the radial coordinate. The resulting equations provide several possibilities for the stated energy conditions. In the particular case of $c=0$, corresponding to the black hole without a surrounding anisotropic fluid, all of the energy conditions are satisfied. Another simple possibility for this to occur corresponds to the value $w_q = 0$. A more general possibility for the energy conditions to be satisfied can be obtained by considering $w_q < 0$, where the other parameters are such that the sum of the terms inside the square brackets is positive in the last two expressions.

\section{Discussion and summary}

In this work, we have provided a new five-dimensional rotating and spherically symmetric Quintessence black hole. To achieve this, we applied the $5D$ version of the Janis-Newman algorithm, which incorporates the Hopf bifurcation, to the static, spherically symmetric quintessence $5D$ black hole solution \cite{Chen:2008ra}. 

Our work focuses, for simplicity, on the case $a \neq 0, b \neq 0, a = b$. By computing the Ricci scalar, we observe that, remarkably, the existence of a physical singularity is only possible for values of the quintessence parameter such that $w_q \in [-0.5, 0]$.

We have noted that there exists a critical value of the quintessence parameter, which, for the parameters used in this work, is $ w_q^{(crit)} \sim -0.5$. For $w_q > w_q^{(crit)}$, there can be up to two horizons, which correspond to an inner horizon and an event horizon, respectively. In this scenario, where $w_q > w_q^{(crit)}$, there is also a critical value of the mass parameter, $M_{cri}$. When $M = M_{crit}$, the inner and outer horizons coincide. This case corresponds to a singular extremal black hole. For $M > M_{crit}$, there is both an inner horizon and a black hole horizon. Thus, this case represents a singular black hole with the presence of both mentioned horizons. For $M < M_{cri}$, there are no horizons, indicating the presence of a rotating $5D$ naked singularity.

On the other hand, for $w_q < w_q^{(crit)}$, there is only one horizon, which corresponds to a cosmological horizon, analogous to what occurs in a de Sitter universe. Therefore, the geometry can be interpreted as a kind of spherically symmetric and rotating universe. For the chosen parameters, in this case, there is no central singularity.

In other words, the variation (increase) of the quintessence parameter causes the geometry to transition from a regular rotating universe surrounded by a cosmological horizon (for $w_q < w_q^{(crit)}$) to a singular rotating geometry (once the parameter reaches a critical value $w_q^{(crit)}$), which can represent (depending on the value of the mass parameter $M$) a naked singularity, a singular extremal black hole, or a singular black hole with an inner horizon and an event horizon. We also notice that, since there is an event horizon, it is surrounded by a surface at radius $r_{s+}$. The region between the event horizon and this last radius, $r_+ < r < r_{s+}$, corresponds to the ergosphere. In this way, we can have both an extremal black hole and a black hole with two horizons (inner and event horizons) with a corresponding region for the ergosphere.

For the study of the null geodesics of photons, we have followed the methodology of reference \cite{Frolov:2003en}. This methodology is based on the Hamilton-Jacobi formalism for studying the equations of motion of null geodesics in $5D$, in Hopf coordinates. In this way, we have determined the radial geodesic equation in the form of a first-order differential equation. In this way, using the obtained radial geodesic equation, we have studied the shape of the shadows of our rotating solution with quintessence. For this, we have followed the recent methodology proposed in reference \cite{Novo:2024wyn}, which, in broad terms, argues that: For the study of shadows in five dimensions, one should adopt the perspective of higher-dimensional beings, whose equivalent of the retina is a volume, meaning it has 3 spatial dimensions. Thus, the authors argue that the $2D$ shadow observed by humans would correspond to cross-sections of this $3D$ shadow.

In the analysis of shadows, we observe that as the magnitude of the quintessence parameter $c$ increases, the size of the shadow decreases. In other words, the presence of quintessence modifies the trajectory of the photons, such that the corresponding photon sphere tends to shrink as the presence of quintessence becomes stronger. Our result differs from the $4D$ study of the Kerr black hole under the influence of quintessence in reference \cite{Singh:2017xle}. In the latter, if we increase the values of the normalization factor $c$, we find that the size of the black hole shadow increases. This difference may be due to the use of the observational device definition mentioned earlier in reference \cite{Novo:2024wyn}.

We have also investigated the shape that the shadow takes as the rotation becomes stronger. In Figure \ref{FigAchatada}, we observe that for a weak rotation parameter ($a = 0.1$ in the example), the shadow is almost a perfect circle. As the rotation increases (for instance, $a = 0.5$ and $a = 0.57$), the distance from the center of the shadow to its highest point along the vertical axis deviates from the radius of the circle, and thus the shadow no longer maintains a circular shape. As the rotation continues to increase, the shadow becomes a closed curve, as shown in the bottom right figure ($a = 0.65$ in the example). In this way, we can observe that, under our methodology and given the influence of quintessence parameters in $5D$, the shape of our shadow differs from the $4D$ quintessence case \cite{Singh:2017xle}, as well as from other models studied in $5D$ \cite{Papnoi:2014aaa,Ahmed:2020jic,Amir:2017slq}.

Furthermore, we have proposed a speculative methodology to test the shadow behavior in five-dimensional scenarios, in light of the constraints provided by the Event Horizon Telescope (EHT) regarding the shadow of the four-dimensional supermassive black hole M87. More specifically, we have formulated the constraints associated with the shadow size and the mass in five dimensions, in an analogous form to the four-dimensional case reported by the EHT for M87. To this end, we have taken the static gravitational potential as a reference. We have computed the values of the shadow radius and the distortion parameter. Our analysis shows that, for fixed values of $w_q$, $a$, and $M$, the ratio in condition~\eqref{RestriccionLL} can satisfy the constraints of the same equation, depending on the value of the quintessence parameter $c$. This suggests that, depending on the strength of the quintessence term, the values of the static gravitational potential in $5D$ can be such that the theoretical predictions remain consistent with the experimental measurements for M87 obtained by the EHT in $4D$ scenarios without quintessence.

The analysis of the shadow radius constraints described above, based on the 5D Newtonian potential, provides a theoretical perspective. However, a more robust measure, since it does not depend on dimensional analysis like the previous one, is the circularity deviation. In this respect, reference \cite{EventHorizonTelescope:2019ths} addressed the investigation of possible constraints on our spacetime geometry based on the measurements of the black hole shadow provided by the EHT M87 collaboration. Specifically, these measurements revealed an asymmetric bright emission ring with a deviation from circularity of $\Delta C \leq 0.1$. To test the circularity deviation, we have followed the methodology proposed in Reference \cite{Bambi:2019tjh} (see also References \cite{Afrin:2021imp,Afrin:2021wlj,Ahmed:2025zdc}). Accordingly, we have calculated the average shadow radius and the deviation from a perfect circular shape. As observed in Table \ref{TablaValores2}, for the quintessence parameter domain $c \in [0, 2]$, and keeping the remaining parameters fixed, the circular deviation $\Delta C$ decreases within the range $[0.0162, 0.0045]$. From this behavior, we can extrapolate that, for the chosen parameter set, $\Delta C \leq 0.1$, which is consistent with the EHT M87 constraints. This analysis suggests that, by employing the method for representing shadows in $5D$ as described in Reference \cite{Novo:2024wyn}, the results satisfy the bound $\Delta C \leq 0.1$ both in the case with quintessence and in the limiting case without quintessence.

The components of the energy-momentum tensor for our solution are shown in detail in Appendix \ref{ComponentesEM}. Due to the non-diagonal structure of the energy-momentum tensor, it becomes difficult to study the energy conditions. In this work, we have used the basis from reference \cite{Aliev:2004ec} to obtain a diagonalized version of the energy-momentum tensor, which allows us to evaluate the energy conditions at the location $\theta=0$.

We have determined numerically that for certain ranges of the radial coordinate, and specific values of the quintessence parameters $w_q$ and $c$, the mass parameter, and the parameter $a$, the energy conditions are satisfied. However, we can state that there are no cases where the NEC and WEC energy conditions are satisfied for all values of the radial coordinate. Despite this, remarkably, the energy density is always positive for the case where our geometry represents a rotating black hole with quintessence.

\section*{appendices}

\appendix

\section{A brief revision of the $5D$ static and spherically symmetric Quintessence black hole} \label{RevisionKiselev}

The five-dimensional static and spherically symmetric Quintessence black hole was studied in reference \cite{Chen:2008ra}. The line element is:
\begin{equation}
  ds^2= -f(r)dt^2+ \frac{dr^2}{f(r)} + r^2 d\Omega_3
\end{equation}
The energy momentum tensor is diagonal such that:
\begin{align}
& T^t_t = T^r_r = - \rho_q \nonumber \\
&T^\phi_\phi=T^\psi_\psi=T^\theta_\theta= \frac{1}{3} \rho_q \left[ 4w_q + 1 \right]
\end{align}

Thus, we can define quintessence's pressure as the average value of the pressure components,

\begin{equation}
    p_q= \langle T^i_j \rangle = \frac{T^r_r+T^\phi_\phi+T^\psi_\psi+T^\theta_\theta}{4}=w_q \rho_q
\end{equation}

For quintessence, we have $-1 < w_q < 0$. The quintessence's energy density is:
\begin{equation}
    \rho_q =- \frac{3 c w_q }{r^{4(w_q + 1)}}
\end{equation}
with $c>0$. 

Thus the lapsus function can be written as: 
\begin{equation}
    f(r) = 1-m(r)/r^2
\end{equation}
where the mass function $m(r)$ is given by equation \eqref{FuncionDeMasa}. Thus
\begin{equation}
    f(r)=1-\frac{2M}{r^2}-\frac{c}{r^{4w_q+2}}
\end{equation}
which in the limit $w_q \to -1$ recovered the $5D$ Schwarzschild Tangherlini dS solution. Furthermore, in the limit $c=0$ recovered the $5D$ Schwarzschild Tangherlini solution.

\section{Review of Geodesic Motion} \label{ApendiceGeodesicas}

To fully solve the system of equations of motion, we need one more constant of motion. To obtain this last one, in the context of black hole shadows, as was mentioned, the Hamilton-Jacobi method is usually employed. The Hamiltonian is such that:
\begin{equation} \label{EqMovHamiltoniano}
- \frac{\partial S}{\partial \lambda} = H = \frac{1}{2} g^{\mu\nu} \frac{\partial S}{\partial x^{\mu}} \frac{\partial S}{\partial x^{\nu}}
\end{equation}
where $S$ is the Jacobian action. A form for $S$, such that the equations of motion are separated into a radial dependence and an angular dependence $\theta$, is provided by references \cite{Frolov:2002xf,Frolov:2003en}:
\begin{equation} \label{AccionJacobbi}
S = \frac{1}{2} \bar{m}^2 \lambda - E t + \Phi \phi + \Psi \psi + S_{\theta} + S_{x}
\end{equation}
where $\bar{m}$ is the mass of the particle and $S_x(x)$, $S_{\theta}(\theta)$ are the functions of $x$ and $\theta$. 

As mentioned previously, this work considers the simplest case where $a=b$. Following \cite{Frolov:2003en,Ahmed:2020jic} and considering that we are interested in a massless photon ($\bar{m} = 0$), using equations \eqref{EqMovHamiltoniano} and \eqref{AccionJacobbi}, we arrive at:
\begin{equation}
   \left ( \frac{\partial S_{\theta}}{\partial \theta} \right )^2 - E^2 a^2 + \frac{1}{\sin^2 \theta} \Phi^2 + \frac{1}{\cos^2 \theta} \Psi^2 = K , 
\end{equation}
\begin{equation}
4 \Delta \left (\frac{\partial S_x}{\partial x} \right)^2  - E^2 x - \frac{m(x)}{\Delta} (x + a^2)^2 \mathcal{E}^2  = -K
\end{equation}
where, in equation \eqref{MathE}, the meaning $\mathcal{E}$ is indicated. Thus, $K$ is the remaining constant of the motion. After some computations, the last two equations can be written in compact form as:
\begin{align}
    &\frac{\partial S_{\theta}}{\partial \theta} = \sigma \sqrt{\Theta}, \\
& \frac{\partial S_{x}}{\partial x} = \sigma \sqrt{X}. \label{EcuacionAccionJacobianaX}
\end{align}
where $\sigma= \pm$ and where:
\begin{equation}
    \Theta(\theta) = E^2 a^2 - \frac{1}{\sin^2 \theta} \Phi^2 - \frac{1}{\cos^2 \theta} \Psi^2 + K
\end{equation}
\begin{equation}
    X=\frac{\mathcal{X}}{4 \Delta^2} \label{EcuacionX}
\end{equation}
where, in equation \eqref{MathX}, the meaning of the last equation is indicated.

\section{Components of the energy-momentum tensor.} \label{ComponentesEM}
\vspace{-1cm}
{\small
\begin{align}
&T_{tt}=\frac{1}{r^3 \Big( a^2 + b^2 + (a^2 - b^2) \cos(2 \theta) + 2 r^2 \Big)^3} \nonumber \\
&\quad \times \Bigg( 2 \Big( a^4 b^2 + a^2 b^4 - a^4 r^2 + 8 a^2 b^2 r^2 - b^4 r^2 + 3 a^2 r^4 + 3 b^2 r^4 + 2 r^6 + m(r) r^2 \Big( a^2 + b^2 + (a^2 - b^2) \cos(2 \theta)  \Bigg. \Bigg. \nonumber \\
&\quad - 2 r^2 \Big) + \Bigg. \Bigg. \cos(2 \theta) \Big( a^4 (b^2 - r^2) + b^2 r^2 \Big( b^2 + 3 r^2 \Big) - a^2 \Big( b^4 + 3 r^4 \Big) \Bigg) \frac{d}{dr} m(r) \Bigg) \Bigg. \nonumber \\
&\quad - 2 r \Big( a^2 + b^2 + (a^2 - b^2) \cos(2 \theta) + 2 r^2 \Big) \times \Big( - m(r) r^2 + (a^2 + r^2) \Big( b^2 + r^2 \Big) \Big) \frac{d^2}{dr^2} m(r)  \nonumber \\ 
&\quad + 2 r^2 \Big( a^2 + b^2 + (a^2 - b^2) \cos(2 \theta) + 2 r^2 \Big) \nonumber \\
&\quad \times \Big( a^2 \cos^2(\theta) - m(r) + r^2 + b^2 \sin^2(\theta) \Big) \Big( 2 \frac{d}{dr} m(r) + r \frac{d^2}{dr^2} m(r) \Big)  
\end{align}
}
{\small
\begin{align}
T_{t \phi}=&\frac{1}{2 r^3 \Big( a^2 + b^2 + (a^2 - b^2) \cos(2 \theta) + 2 r^2 \Big)^3} \nonumber \\
\times & a \sin^2(\theta) \Bigg( -4 \Big( m(r) r^2 \Big( a^2 + b^2 + (a^2 - b^2) \cos(2 \theta) - 2 r^2\Big) \nonumber \\
+& \Big( a^2 + r^2 \Big) \Big( b^4 + 9 b^2 r^2 + 6 r^4 + a^2 (b^2 - r^2) + (a^2 - b^2) \cos(2 \theta) (b^2 - r^2) \Big) \Big) \frac{d}{dr} m(r) \nonumber \\
+& 4 r \Big( a^2 + b^2 + (a^2 - b^2) \cos(2 \theta) + 2 r^2 \Big) \Big( - m(r) r^2 + \Big( a^2 + r^2 \Big) \Big( b^2 + r^2 \Big) \Big) \frac{d^2}{dr^2} m(r) \nonumber \\
+& 4 m(r) r^2 \Big( a^2 + b^2 + (a^2 - b^2) \cos(2 \theta) + 2 r^2 \Big) \times \Big( 2 \frac{d}{dr} m(r) + r \frac{d^2}{dr^2} m(r) \Big) \Bigg )
\end{align}
}
{\small
\begin{align}
T_{t \psi}=&\frac{1}{2 r^3 \Big( a^2 + b^2 + (a^2 - b^2) \cos(2 \theta) + 2 r^2 \Big)^3} \nonumber \\
\times & b \cos^2(\theta) \Bigg( -4 \Big( m(r) r^2 \Big( a^2 + b^2 + (a^2 - b^2) \cos(2 \theta) - 2 r^2 \Big) \nonumber \\
+& \Big( b^2 + r^2 \Big) \Big( a^4 - b^2 r^2 + 6 r^4 + (a^2 - b^2) \cos(2 \theta) (a^2 - r^2) + a^2 \Big( b^2 + 9 r^2 \Big) \Big) \Bigg) \frac{d}{dr} m(r)  \nonumber \\
+ & 4 r \Big( a^2 + b^2 + (a^2 - b^2) \cos(2 \theta) + 2 r^2 \Big)\Big( - m(r) r^2 + \Big( a^2 + r^2 \Big) \Big( b^2 + r^2 \Big) \Big) \frac{d^2}{dr^2} m(r) \nonumber \\
+& 4 m(r) r^2 \Big( a^2 + b^2 + (a^2 - b^2) \cos(2 \theta) + 2 r^2 \Big) \Big( 2 \frac{d}{dr} m(r) + r \frac{d^2}{dr^2} m(r) \Big) \Bigg)
\end{align}
}
\begin{align}
 T_{rr} = &\Bigg( \frac{r \left( a^2 + b^2 + \left( a^2 - b^2 \right) \cos\left( 2 \theta \right) + 6 r^2 \right) \frac{\partial m}{\partial r}}{2 \left( a^2 + b^2 + \left( a^2 - b^2 \right) \cos\left( 2 \theta \right) + 2 r^2 \right) \left( -m\left( r \right) r^2 + \left( a^2 + r^2 \right) \left( b^2 + r^2 \right) \right)} \Bigg)
\end{align}
\begin{equation}
T_{\theta \theta}= -\frac{2 \left( a^2 \cos^2(\theta) + b^2 \sin^2(\theta) \right) \frac{d}{dr} m(r) + r \left( a^2 \cos^2(\theta) + r^2 + b^2 \sin^2(\theta) \right) m''(r)}{r \left( a^2 + b^2 + (a^2 - b^2) \cos(2\theta) + 2 r^2 \right)}
\end{equation}
{\small
\begin{align}
T_{\phi \phi}=&\frac{1}{2 r^3 \Big( a^2 + b^2 + (a^2 - b^2) \cos(2\theta) + 2 r^2 \Big)^{\!3}} \sin^2(\theta) \Bigg[ \big( a^2 + r^2 \big) \Bigg( a^4 b^2 + 3 a^2 b^4 - a^4 r^2 \nonumber \\
+ & 19 a^2 b^2 r^2 + 6 b^4 r^2 + 20 a^2 r^4 + 16 b^2 r^4 + 16 r^6 - (a^2 - b^2) \cos(4\theta) \Big( 2 b^2 r^2 + a^2 (b^2 - r^2) \Big) \Bigg) \nonumber \\
 -& 4 \cos(2\theta) \Big( 2 b^2 r^2 \big( b^2 + 2 r^2 \big) + a^2 \big( b^4 + 4 b^2 r^2 + r^4 \big) \Big) \Bigg] \nonumber \\
+& 4 a^2 m(r) r^2 \Big( a^2 + b^2 + (a^2 - b^2) \cos(2\theta) - 2 r^2 \Big) \sin^2(\theta) m'(r) \nonumber \\
-& 4 a^2 r \Bigg[ a^2 + b^2 + (a^2 - b^2) \cos(2\theta) + 2 r^2 \Bigg] \Bigg[ - m(r) r^2 + (a^2 + r^2) (b^2 + r^2) \Bigg] \nonumber \\
& \sin^2(\theta) m''(r) \Bigg) - r^2 \Big( a^2 + b^2 + (a^2 - b^2) \cos(2\theta) + 2 r^2 \Big)^2 \Bigg( a^2 + r^2 \nonumber \\
+& \frac{a^2 m(r) \sin^2(\theta)}{a^2 \cos^2(\theta) + r^2 + b^2 \sin^2(\theta)} \Bigg) \Big( 4 m'(r) + 2 r m''(r) \Big) \Bigg)
\end{align}
}
{\small
\begin{align}
& T_{\phi \psi} = \frac{1}{2 r^3 \Bigg( a^2 + b^2 + (a^2 - b^2) \cos(2\theta) + 2 r^2 \Bigg)^3} \nonumber \\
& \quad \cdot a b \Bigg( - 4 \cos^2(\theta) m(r) r^2 \Bigg( a^2 + b^2 + (a^2 - b^2) \cos(2\theta) + 2 r^2 \Bigg) \sin^2(\theta) \Bigg( 2 \frac{d}{dr} m(r) \nonumber \\
& \quad + r \cdot m''(r) \Bigg) + \sin^2(2\theta) \Bigg( m(r) r^2 \Bigg( a^2 + b^2 + (a^2 - b^2) \cos(2\theta) - 2 r^2 \Bigg) \nonumber \\
& \quad + (a^2 + r^2) (b^2 + r^2) \Bigg( a^2 + b^2 + (a^2 - b^2) \cos(2\theta) + 10 r^2 \Bigg) \Bigg) \frac{d}{dr} m(r) \nonumber \\
& \quad - r \Bigg( a^2 + b^2 + (a^2 - b^2) \cos(2\theta) + 2 r^2 \Bigg) \Bigg( - m(r) r^2 + (a^2 + r^2) (b^2 + r^2) \Bigg) m''(r) \Bigg) \Bigg) \Bigg)  
 \end{align}
 }
{\small
 \begin{align}
&T_{\psi \psi}= \frac{\cos^2(\theta)}{2 r^3 \Bigg( a^2 + b^2 + (a^2 - b^2) \cos(2\theta) + 2 r^2 \Bigg)^{\!3}}  \Bigg( \Bigg( 4 b^2 \cos^2(\theta) m(r) r^2 \Bigg( a^2 + b^2 + (a^2 - b^2) \cos(2\theta)  \nonumber \\
& - 2 r^2 \Bigg) + (b^2 + r^2) \Bigg( 3 a^4 b^2 + a^2 b^4 + 6 a^4 r^2 + 19 a^2 b^2 r^2 - b^4 r^2 + 16 a^2 r^4 + 20 b^2 r^4 + 16 r^6 \nonumber \\
& \quad + (a^2 - b^2) \cos(4\theta) \Bigg( - b^2 r^2 + a^2 \left( b^2 + 2 r^2 \right) \Bigg) + 4 \cos(2\theta) \Bigg( b^2 r^4 + 4 a^2 r^2 \left( b^2 + r^2 \right) \nonumber \\
& \quad + a^4 \left( b^2 + 2 r^2 \right) \Bigg) \Bigg) m'(r) \nonumber \\
& \quad + 4 b^2 \cos^2(\theta) r \Bigg( - a^2 - b^2 + \left( - a^2 + b^2 \right) \cos(2\theta) - 2 r^2 \Bigg) \Bigg( - m(r) r^2 + \left( a^2 + r^2 \right) \left( b^2 + r^2 \right) \Bigg) \nonumber \\
& \quad \cdot m''(r) \Bigg) - r^2 \Bigg( a^2 + b^2 + (a^2 - b^2) \cos(2\theta) + 2 r^2 \Bigg)^2 \Bigg( b^2 + r^2 + \frac{b^2 \cos^2(\theta) m(r)}{a^2 \cos^2(\theta) + r^2 + b^2 \sin^2(\theta)} \Bigg) \nonumber \\
& \quad \cdot \Bigg( 4 m'(r) + 2 r m''(r) \Bigg) \Bigg)
\end{align}
}
\section*{Acknowledgements}
Milko Estrada is funded by ANID , FONDECYT de Iniciaci\'on en Investigación 2023, Folio 11230247. MSC is partially funded by Conselho Nacional de Desenvolvimento Científico e Tecnológico (CNPq) under the grant number 315926/2021-0. L.C.N.S. would like to thank FAPESC for financial support under grant number 735/2024.

\bibliography{mybib.bib}

\end{document}